\documentclass[11pt,aps,prd,amsmath,amssymb,superscriptaddress,floatfix,nofootinbib,notitlepage]{revtex4-1}
\usepackage{amsmath,amssymb} \usepackage{epsfig} \usepackage{color}
\usepackage{bm} \usepackage{epstopdf}
\usepackage{rotating}
\usepackage[left=2.5cm,right=2.5cm,top=2cm,bottom=2cm]{geometry}
\usepackage[colorlinks=true,hyperfootnotes=false]{hyperref}
\makeatletter
\def\slashchar#1{{\mathpalette\c@ncel{#1}}} 
\makeatother
\def\vsl{\slashchar{v}}

\newcommand{\X}{X(3872)}
\newcommand{\Tr}{\text{Tr}}
\newcommand{\mbare}{\stackrel{\circ}{m}_{c\bar c}}
\newcommand{\mbareuno}{\stackrel{\circ}{m}_{\chi_{c1}}}
\newcommand{\mbaredos}{\stackrel{\circ}{m}_{\chi_{c2}}}
\newcommand{\mbareunob}{\stackrel{\circ}{m}_{\chi_{b1}}}
\newcommand{\mbaredosb}{\stackrel{\circ}{m}_{\chi_{b2}}}

\begin{document}

\title{Quarkonium Contribution to Meson Molecules}

\author{E. Cincioglu}
\email{elif.cincioglu@gmail.com}
\affiliation{Department of Physics Engineering, Ankara University, Ankara, Turkey}
\author{J. Nieves}
\affiliation{Instituto de F\'\i sica Corpuscular (IFIC), Centro Mixto
CSIC-Universidad de Valencia, Institutos de Investigaci\'on de
Paterna, E-46071 Valencia, Spain}
\author{A. Ozpineci}
\affiliation{Department of Physics, Middle East Technical University, Ankara, Turkey}
\author{A. U. Yilmazer}
\affiliation{Department of Physics Engineering, Ankara University, Ankara, Turkey}

\date{\today}

\begin{abstract}

Starting from a molecular picture for the $X(3872)$ resonance, this
state and its $J^{PC}=2^{++}$ heavy quark spin symmetry partner
$[X_2(4012)]$ are analyzed within a model which incorporates possible
mixings with $2P$ charmonium ($c\bar c$) states.  Since it is
reasonable to expect the bare $\chi_{c1}(2P)$ to be located above the
$D\bar D^*$ threshold, but relatively close to it, the presence of the
charmonium state provides an effective attraction that will
contribute to bind the $X(3872)$, but it will not appear in the
$2^{++}$ sector. Indeed in this latter sector, the $\chi_{c2}(2P)$
should provide an effective small repulsion, because it is placed well below the
$D^*\bar D^*$ threshold. We show how the $1^{++}$ and
$2^{++}$ bare charmonium poles are modified due to the $D^{(*)}\bar
D^{(*)}$ loop effects, and the first one is moved to the complex
plane.  The meson loops produce, besides some shifts in the masses of
the charmonia, a finite width for the $1^{++}$ dressed charmonium
state.  On the other hand, the $X(3872)$ and $X_2(4012)$ start
developing some charmonium content, which is estimated by means of the
compositeness Weinberg sum-rule. It turns out that in the heavy quark
limit, there is only one coupling between the $2P$ charmonia and the
$D^{(*)}\bar D^{(*)}$ pairs. We also show that for reasonable values
of this coupling, leading to $X(3872)$ molecular probabilities of
around 70-90\%, the $X_2$ resonance destabilizes and disappears from
the spectrum, becoming either a virtual state or being located deep
into the complex plane, with decreasing influence in the $D^{*}\bar
D^{*}$ scattering line. Moreover, we also discuss how around 10-30\%
charmonium probability in the $X(3872)$ might explain the ratio of
radiative decays of this resonance into $\psi(2S)\gamma$ and $J/\psi
\gamma$. Finally, we qualitatively discuss within this scheme, the
hidden bottom flavor sector, paying a special attention to the
implications for the $X_b$ and $X_{b2}$ states, heavy quark spin
flavor partners of the $X(3872)$.

\end{abstract}


\maketitle

\section{Introduction}

The $X(3872)$ state was first observed by the Belle collaboration
\cite{Choi:2003ue} in the $B^{\pm}\to J/\psi\pi^{+}\pi^{-}K^{\pm}$
channel as a narrow peak and was confirmed by various other
experiments
\cite{Acosta:2003zx,Abazov:2004kp,Aubert:2004ns,Aaij:2011sn}. The
averaged mass of $X(3872)$ is $3871.69\pm 0.17$ MeV, which is only $0.16$
MeV below the $D^0\bar{D}^{*0}$ threshold and the full width is less
than $1.2$ MeV \cite{Agashe:2014kda}. In addition, the LHCb experiment determined its $J^{PC}$ quantum
numbers as $1^{++}$ \cite{Aaij:2013zoa}. The properties of $X(3872)$
turned out to be difficult to reconcile with a $c \bar c$ state in a
quark potential model picture \cite{Barnes:2003vb, Suzuki:2005ha}. Alternative
theoretical models have been proposed to understand its structure. One
of the popular descriptions of $X(3872)$ is as a molecular state
consisting of a $D$ and a $\bar D^*$
\cite{Swanson:2003tb, Voloshin:2004mh, Braaten:2005ai, 
Gamermann:2007fi, Liu:2007bf,Liu:2008fh,Dong:2008gb,Gamermann:2009uq}.

One of the puzzling observations about $X(3872)$ is the ratio of its
decays into final states with isospin-0 and isospin-1. The ratio of
the decay fractions of $X(3872)$ into  $J/\psi\pi^{+}\pi^{-}$ and
into  $J/\psi\pi^{+}\pi^{-}\pi^{0}$ final states was first measured by
Belle \cite{Abe:2005ix} to be:
\begin{equation}
\frac{Br(J/\psi\pi^{+}\pi^{-}\pi^{0})}{Br(J/\psi\pi^{+}\pi^{-})}=1.0\pm 0.4\pm 0.3.
\end{equation}
For the same ratio, BABAR has obtained $1.0\pm 0.8\pm 0.3$
\cite{delAmoSanchez:2010jr}. Later Belle announced  the updated results of the
measurements for the reaction $J/\psi\pi^{+}\pi^{-}\pi^{0}$, and thus the
accepted combined result from Belle  and BABAR is $0.8\pm
0.3$~\cite{Choi:2011fc}.  The decays into final states with two and
three pions proceed through  virtual $\rho$ and $\omega$ mesons, respectively.  Considering
the phase space differences between the $\rho$ and $\omega$ mesons,
the production amplitude ratio is found to be~\cite{Hanhart:2011tn}
\begin{equation}
\Big{|}\frac{A(J/\psi\rho)}{A(J/\psi\omega)}\Big{|}=0.26\pm 0.07
\end{equation}
Such a large isospin violation arises
naturally in the molecular picture due to the mass difference between
the $D^{0}\bar{D}^{*0}$ and $D^{+}D^{*-}$ components in the $X(3872)$
wave function \cite{Gamermann:2009fv,Gamermann:2009uq}, and the
remarkable proximity of the resonance to the $D^0\bar D^{0*}$ threshold.

Other interesting $X(3872)$ measurements are its radiative
decays. The ratio of the branching fractions into final states with a
photon and a $J/\psi$ or a $\psi(2S)$ has been measured as
\cite{Aubert:2008ae,Aaij:2014ala}:
\begin{equation}
R_{\psi\gamma}=\frac{Br(X\to\psi(2S)\gamma)}{Br(X\to J/ \psi
  \gamma)}=2.46\pm 0.64\pm 0.29 \label{eq:Rexprad}
\end{equation}
One of the first works, where the radiative decays of the $X(3872)$ was
studied within an effective field theory framework, was carried out in
\cite{Mehen:2011ds}. There, the $X(3872)\to\psi(2S)\gamma$ reaction was
studied and some qualitative conclusions were drawn. It was argued that
the decay should receive contribution from long-distance 
physics, involving the propagation of intermediate heavy charm mesons
($D^0\bar D^{*0}-hc)$, and short-distance dynamics, whose contribution
is encoded in a contact
operator.  The  $\chi_{c1}(2P)$
state contributed to the latter operator, through $D \bar D^* \to
\chi_{c1}(2P) \to \psi(2S)\gamma$.  The relative importance of
these two types of contributions was unknown, though it was shown in
\cite{Mehen:2011ds} that the angular distributions of the decay products can
be used to distinguish between them.

There were claims~\cite{Swanson:2004pp} that within the molecular
picture, such a large ratio can not be naturally explained.  This
ratio can be however accommodated assuming that there is a charmonium
admixture in the molecular state~\cite{Badalian:2012jz,  Wang:2010ej, Eichten:2005ga, Dong:2009uf}. Thus for instance, an
enhanced decay of the $X(3872)$ into $\psi(2S) \gamma$ compared to
$J/\psi \gamma $, and fully compatible with a predominantly molecular
nature of $X(3872)$ was found in Ref.~\cite{Dong:2009uf}, where a
phenomenological study allowing for both a molecular as well as a
compact component of the $X(3872)$ was carried out. Actually, an
admixture of 5--12\% of a $\bar c c$ component was sufficient to
explain the data~\cite{Dong:2009uf}.  This charmonium admixture is
also favored by the production rate of $X(3872)$ in the $p\bar{p}$
collisions which is about $1/20$ of the rate of $\psi(2S)$. This
production rate can easily be explained if one assumes that the
$c\bar{c}$ component of $X(3872)$ is approximately $5\%$
\cite{Takizawa:2012hy}.

The validity of the claim of Ref.~\cite{Dong:2009uf} was critically
reviewed in Ref.~\cite{Guo:2014taa} from an effective field theory
(EFT) point of view. There, it was concluded, contrary to
earlier claims, that radiative decays do not allow one to draw conclusions
on the nature of $X(3872)$. Actually,  the
findings of Ref.~\cite{Dong:2009uf}  were qualitatively confirmed, and
in addition it was pointed out that the observed ratio is not in conflict with a
predominantly molecular nature of the $X(3872)$. The study
of Ref.~\cite{Guo:2014taa} suggests that for  radiative decays
of the  $X(3872)$, short-range contributions are of similar importance as their long-range counter parts

 In the heavy quark limit, an EFT to describe the $X(3872)$ and also
 other possible $D^{(*)} \bar D^{(*)}$ molecules has been proposed in
 \cite{Nieves:2012tt,HidalgoDuque:2012pq}. At very low energies, the
 leading order~(LO) interaction between the $D^{(*)}\bar D^{(*)}$
 mesons can be described just in terms of contact-range potentials,
 which are constrained by heavy quark spin symmetry (HQSS). Pion
 exchange and particle coupled-channel\footnote{We do not refer to
   charge channels, but rather to the mixing among the $D\bar D$,
   $D\bar D^*$, $D^*\bar D^*$ pairs in a given $IJ$ (isospin and spin)
   sector.} effects are conjectured to be sub-leading, and they are
 not considered at LO, within the scheme advocated in
 ~\cite{Nieves:2012tt,Valderrama:2012jv}, where it is assumed that
 HQSS is respected in the interactions, but broken by the heavy-light
 meson masses. This scheme, in principle, should make sense for
 loosely bound molecules, as their binding is smaller than the meson
 mass splittings, and it requires the use of ultraviolet (UV)
 regulators sufficiently small to prevent violations of HQSS. In
 \cite{Nieves:2012tt,Valderrama:2012jv}, it is argued on general
 grounds that expected coupled channels effects should be suppressed
 by the square of the ratio of the light scale over the coupled
 channel momentum scale, which in the charm sector is around 500-700
 MeV. Moreover, the consideration of coupled channels induced a strong
 dependence on the UV
 regulator~\cite{Nieves:2012tt,Valderrama:2012jv}, which would require
 the inclusion of additional counter-terms to compensate it,
 increasing thus the number of undetermined low energy constants
 (LECs).

Within the molecular description of the $X(3872)$, among others, the
existence of a $X_{2}$ [$J^{PC} = 2^{++}$] $S$-wave $D^{*}\bar{D}^{*}$
bound state was predicted in the EFT approach of
Refs.~\cite{Nieves:2012tt,HidalgoDuque:2012pq}, with a binding energy
similar to that of the $X(3872)$ ($M_{X_2} - M_{X(3872)} \approx
M_{D^*} - M_{D} \approx 140~\text{MeV}$).
Both the $X(3872)$ and the $X_2$ would have partners in the bottom
sector~\cite{Guo:2013sya}\footnote{In Ref.~\cite{Guo:2013sya}, the
  bottom and charm sectors are connected by assuming the bare
  couplings in the four-meson interaction Lagrangian to be independent of the
  heavy quark mass.}, which we will call $X_b$ and $X_{b2}$,
respectively, with masses approximately related by $M_{X_{b2}} -
M_{X_b} \approx M_{B^*} - M_{B} \approx 46~\text{MeV}$. States with
$2^{++}$ quantum numbers exist as well as spin partners of the
$1^{++}$ states in the spectra of the conventional heavy quarkonia and
tetraquarks. However, the mass splittings would only accidentally be
the same as the fine splitting between the vector and pseudoscalar
charmed mesons.

Some exotic hidden charm sectors  have been also recently
studied on the lattice~\cite{Liu:2012ze,Prelovsek:2013cra,Prelovsek:2013xba,Prelovsek:2014swa,Padmanath:2015era},
and evidence for the $X(3872)$ from $D\bar D^*$ scattering on the
lattice has been found~\cite{Prelovsek:2013cra}. The
$2^{++}$ sector has not been exhaustively addressed yet, though a
state with these quantum numbers and a mass of $(m_{\eta_c}+1041\pm 12)
~\rm{MeV}$= $(4025 \pm 12)~\text{MeV}$, close to the value predicted in
Refs.~\cite{Nieves:2012tt,HidalgoDuque:2012pq}, was reported in
Ref.~\cite{Liu:2012ze}, though the calculations were performed with a 
pion mass $\simeq 400$ MeV.  There
exists also a feasibility study~\cite{Albaladejo:2013aka} of future
lattice QCD (LQCD)
simulations, where the EFT approach of
Refs.~\cite{Nieves:2012tt,HidalgoDuque:2012pq} was formulated in a
finite box.

Despite the theoretical predictions on the existence of the $X_2$,
$X_b$ and $X_{b2}$ states, none of these hypothetical particles has
been observed so far. This negative result could be because the
current experiments are not
yet sensitive enough  or due to the non-existence of these states. Nevertheless, they are being and will be
searched for in current and future experiments such as BESIII, LHCb,
CMS, Belle-II and PANDA.

The HQSS EFT approach of
Refs.~\cite{Nieves:2012tt,HidalgoDuque:2012pq} does not consider possible
mixings  between molecular heavy-light meson-antimeson and 
quarkonium states. However in the LQCD simulation carried out in
Ref.~\cite{Prelovsek:2013cra}, it was needed to consider both $c\bar
c-$charmonium and $D\bar D^*-$molecular type interpolating fields to
find a signature\footnote{There, it was also found that the effect of
  the $J/\psi \omega$ channel is irrelevant for the dynamics of the
  $X(3872)$. In that exploratory work, isospin breaking effects were
  not considered, and thus the resonance reported in
  ~\cite{Prelovsek:2013cra} was purely isoscalar.} of the $X(3872)$.
As discussed above, the presence of $c\bar c$ components in the
$X(3872)$ seems to be also required to explain the experimental value
for the ratio of radiative branching fractions $R_{\psi\gamma}$,
quoted in Eq.~\eqref{eq:Rexprad}. Moreover, the charmonium
$\chi_{c1}(2P)$ state, that would have the same quantum numbers
$1^{++}$ as the $X(3872)$, has not been found yet. 

The charmonium admixture in a molecular picture of the $X(3872)$ has
been studied, among others, in Refs.~\cite{Dong:2009uf, Ortega:2010qq,
  Takizawa:2012hy}.  In Ref.~\cite{Takizawa:2012hy}, direct
interactions between the $D$ and $\bar D^*$ mesons are supposed to
play a marginal role, being the coupling to the $c\bar c$ core more
important in creating the $X(3872)$ than the direct $D \bar D^*$
attraction, which is assumed to be independent of the isospin as well
as of the heavy quark masses. The strength of the $D \bar D^*$
attraction is estimated to be barely strong enough to make a
weakly bound state by looking at the experimental masses of the isovector $Z_b(10610)$ and $Z_b(10650)$ resonances,
placed very close to the $B \bar B^{*}$ and $B^* \bar B^{*}$
thresholds, respectively. This rationale might be incorrect since the
$D \bar D^*$ interaction for isospin 1 is suppressed in the large
$N_C$ (number of colors) counting with respect to that in the isoscalar sector. A
non-relativistic constituent quark model is used in
Ref.~\cite{Ortega:2010qq}, and two and four-quark configurations are
coupled using the phenomenological $^3P_0$ model. Finally, the
approach of Ref.~\cite{Dong:2009uf} is based on phenomenological
hadron Lagrangians and the quark model results of
Ref.~\cite{Swanson:2003tb}, where it is proposed that the $X(3872)$ is a 
$D^0\bar D^{*0}$ hadronic resonance stabilized by admixtures of
$\omega J/\psi$ and $\rho J/\psi$.  These works neither made use of HQSS, nor
address the dynamics of possible heavy quark spin-flavor partners of
the $X(3872)$ states. There exist however, some preliminary
results~\cite{Entem:2016ojz}, obtained within the quark model of
Ref.~\cite{Ortega:2010qq}, about the possible existence of heavy quark
spin-flavor partners of the $X(3872)$. 

It is therefore  timely and relevant to extend the HQSS model
of Refs.~\cite{Nieves:2012tt,HidalgoDuque:2012pq} to incorporate 
quarkonium degrees of freedom, and their possible mixings with the molecular
components. This is the  objective of the present work, where we will
make use of HQSS and the experimental ratio $R_{\psi\gamma}$ to
constrain the interaction of the $D^{(*)}\bar D^{(*)}$
pairs with the $2P$ charmonia. (Due to the closeness of their masses, the charmonium
admixture in the $X(3872)$ should correspond to the $2P$ $c\bar c$
states.) We will also study the effects of non-zero quarkonium components  on the predictions for the  $X_2$, $X_b$
and $X_{b2}$ states. We will show that even small mixings between 
charmonium and molecular components in the $X_2$ state  might explain 
why it has not been observed yet. In the hidden bottom sector, however, we
will see how despite the  changes induced by the
quarkonium admixtures, it might be reasonable to expect that both $X_b$ and $X_{b2}$
resonances should be real QCD states, which  might be observed in the
short future. 

In Ref.~\cite{Baru:2016iwj} and working in the strict heavy-quark
limit, the degeneracy of the $X_2$ and $X(3872)$ states was confirmed
as a robust result with respect to the inclusion of the one-pion
exchange interaction between the $D^{(*)}$ mesons. There, it is shown
that this is true if all relevant partial waves as well as particle
channels which are coupled via the pion-exchange potential are taken
into account. Beyond the heavy quark limit and treating
non-perturbatively the pions, in \cite{Baru:2016iwj} it is predicted,
contrary to the findings of Refs.~\cite{Nieves:2012tt,
  Albaladejo:2013aka} obtained with perturbative pions, a significant shift of
the $X_2$ mass and width of the order of 50 MeV. The increase of the
$X_2$ binding energy is only viewed in \cite{Baru:2016iwj} as a
qualitative result. However, the conclusion on the broadening of the
$X_2$ is claimed in that work as a reliable prediction, since it is
argued there that is related to unitarity.  We think these findings
have to be interpreted with some caution. First, one should bear in
mind that the UV cutoffs used in \cite{Baru:2016iwj} are much larger
(around a factor of 2) than those considered in the approach of
Refs. \cite{Nieves:2012tt,Albaladejo:2013aka}. Thus some extra HQSS
breaking corrections, beyond those due to the heavy-light meson
masses, are accounted for in \cite{Baru:2016iwj}, which have indeed
relevance in the numerical results. Such corrections are largely cut
in Refs. \cite{Nieves:2012tt,Albaladejo:2013aka}, and it is not clear
whether they should be considered or not, and given the poor experimental status, it is
difficult to disentangle among both approaches. Second, the hadronic
$D$-wave $X_{2} \to D \bar{D}$ and $X_{2} \to D \bar{D}^{*}$ two-body
decays, driven via one pion exchange, were predicted in
\cite{Albaladejo:2013aka} to be smaller altogether than 5 MeV. There,
large contributions from highly virtual pions carrying large momenta,
which lay outside the range of applicability of the EFT as proposed in
Refs. \cite{Nieves:2012tt,Albaladejo:2013aka} were found. Such
contributions were further suppressed in \cite{Albaladejo:2013aka} by
including an extra form factor in the vertices involving virtual
pions. As can be seen in Table I of this latter reference, $X_2$
widths as large as 30 MeV could be obtained without including this
extra form-factor. Thus, it is not surprising that values of around 50
MeV were found in \cite{Baru:2016iwj} for the width of this resonance
since, as mentioned above, there much larger UV regulators were used.

In what follows, we will use the EFT as conjectured in
Refs. \cite{Nieves:2012tt,Albaladejo:2013aka} and will neglect pion
exchange and coupled channel effects in this preliminary study of the
interplay between quark and meson-molecular degrees of freedom.
However, one should consider also the possibility of a broad
$X_2$ state from a purely molecular picture, as found in the approach
pursued in Ref.~\cite{Baru:2016iwj}, which nevertheless would be also affected
by the consideration of the quark degrees of freedom discussed in the
present work.

This paper is organized as follows. In Sect.~\ref{sec:effLa}, and within a
framework suited to implement HQSS constraints, we introduce the
heavy quark fields and their interactions, including those responsible
for the mixing between meson--meson pairs and $P-$wave quarkonium
states. Also in this section, the $2P \to 1S, 2S$ charmonium radiative
transitions are studied (Subsect.~\ref{sec:char-rad}). In the next
section, Sect.~\ref{sec:unitarized}, the procedure used to obtain
unitarized amplitudes, from the HQSS interactions introduced in the
previous section, is described.  A special attention
(Subsect.~\ref{sec:LSE}) is paid to a non-perturbative re-summation
based on the solution of a renormalized Lippmann--Schwinger equation
(LSE). In Sect.~\ref{sec:generalidades}, some general properties of
the poles of the unitarized amplitudes and the compositeness
condition, which will serve us to quantify the importance of the
molecular components in the resonances, are discussed. Specific
formulas for the two-channel problem relevant to study the $1^{++}$
and $2^{++}$ hidden charm or bottom meson molecules are given in the
first part of Sect.~\ref{sec:char-1++2++}. Numerical results on the
influence of the quarkonium components in the properties of the
$X(3872), X_2(4012),X_b$ and $X_{b2}$ meson molecules are presented and discussed in
Subsects.~\ref{sec:X3872}, \ref{sec:2++charm} and
\ref{sec:bottom}. Within the Subsect.~\ref{sec:X3872}, a numerical
study of the $X(3872)\to J/\psi \gamma$ and $\psi(2S)\gamma$ transitions,
based on Subsect.~\ref{sec:char-rad} and Ref.~~\cite{Guo:2014taa}, is
presented and used to constrain the charmonium content in the $X(3872)$. The most relevant findings of this work are
summarized in Sect.~\ref{sec:concl}, and finally, the properties of the $1^{++}$ and
$2^{++}$ hidden charm and bottom poles discussed in the previous
sections, but calculated with a different UV regulator
are collected in Appendix~\ref{sec:appendix}.

\section{LO effective  Lagrangians}
\label{sec:effLa}
\subsection{HQSS fields}

We use the matrix field $H^{(Q)}$ [$H^{(\bar Q)}$] to
describe the combined isospin doublet of pseudoscalar heavy-mesons
$P^{(Q)}_a=(Q\bar u,Q\bar d)$ [$P^{(\bar Q)a}=(u \bar Q,d \bar Q )^t$] fields and
their vector HQSS partners $P^{*(Q)}_a$ [$P^{*(\bar Q)a}$] (see for
example \cite{Grinstein:1992qt}),
\begin{eqnarray}
H_a^{(Q)} &=& \frac{1+\vsl}2 \left (P_{a\mu}^{* (Q)}\gamma^\mu -
P_a^{(Q)}\gamma_5 \right), \qquad v\cdot P_{a}^{* (Q)} = 0,  \nonumber \\
H^{(\bar Q)a} &=&  \left (P_{\mu}^{* (\bar Q)a}\gamma^\mu -
P^{(\bar Q)a}\gamma_5 \right) \frac{1-\vsl}2 , \qquad v\cdot P^{*
  (\bar Q)a} = 0.
\end{eqnarray}
The matrix field $H^{c}$ [$H^{\bar c}$] annihilates $P$ [$\bar P$]
and $P^*$ [$\bar P^*$] mesons with a definite velocity $v$. Under a
parity transformation we have
\begin{equation}
H^{(Q,\bar Q)}(x^0,\vec{x}) \to \gamma^0  H^{(Q,\bar Q)}(x^0,-\vec{x})\gamma^0, \qquad v^\mu \to v_\mu
\end{equation}
The field
$H_a^{(Q)}$ [$H^{(\bar Q)a}$] transforms as a $(2,\bar 2)$ [$(\bar
  2,2)$] under the heavy spin $\otimes $ SU(2)$_V$ isospin
symmetry~\cite{Grinstein:1992qt}, this is to say:
\begin{equation}
H_a^{(Q)} \to S_Q \left( H^{(Q)} U ^\dagger\right)_a, \qquad
H^{(\bar Q) a} \to \left(U  H^{(\bar Q)}\right)^a S^\dagger_{\bar Q}.
\end{equation}
Their hermitian conjugate fields are defined by:
\begin{equation}
\bar H^{(Q)a} =\gamma^0 [H_a^{(Q)}]^\dagger \gamma^0, \qquad
\bar H_a^{(\bar Q)} =\gamma^0  [H^{(\bar Q)a}]^\dagger \gamma^0 ,
\end{equation}
and transform as~\cite{Grinstein:1992qt}:
\begin{equation}
\bar H^{(Q)a} \to  \left( U \bar H^{(Q)} \right)^a S^\dagger_Q , \qquad
\bar H^{(\bar Q)}_a \to S_{\bar Q}\left(\bar H^{(\bar Q)} U^\dagger \right)_a .
\end{equation}
The definition for $H_a^{(\bar Q)}$ also
specifies our convention for charge conjugation, which is $\mathcal{C}P_a^{(Q)}
\mathcal{C}^{-1} = P^{(\bar Q) a} $ and $\mathcal{C}P_{a\mu}^{*(Q)}\mathcal{C}^{-1}
= -P_\mu^{*(\bar Q) a} $, and thus it follows
\begin{equation}
\mathcal{C}H_a^{(Q)} \mathcal{C}^{-1} = c\, H^{(\bar Q)at}\, c^{-1}, \qquad
\mathcal{C}\bar H^{(Q)a}\mathcal{C}^{-1} = c\, \bar H_a^{(\bar Q)t}\, c^{-1}
\end{equation}
with $c$ the Dirac space charge conjugation matrix satisfying $c\gamma_\mu
c^{-1}=-\gamma_\mu^t$, and $t$ denotes the matrix transpose operation.

A heavy quark--antiquark bound state, characterized by the radial
number $n$, the orbital angular momentum $l$, the spin $s$ and the
total angular momentum $J$, is denoted by $n\, ^{2s+1}l_J$. Parity and
charge conjugation are given by $P=(-1)^{l+1}$, $C=(-1)^{l+s}$.  If
spin dependent interactions are neglected it is natural to describe
the spin singlet $n\,^1l_{J=l} $ and the spin triplet $n\, ^3l_{J=l-1,l,l+1}$ by means
of a single multiplet $\hat J(n, l)$. For $l=0$, when the triplet
$s= 1$ collapses into a single state with total angular momentum
$j=1$, this is readily realized by adopting the
description~\cite{Jenkins:1992nb}
\begin{equation}
\hat J = \frac{1+\vsl}{2}\left ( \psi_\mu \gamma^\mu -\gamma_5 \eta \right)\frac{1-\vsl}2
\end{equation}
Here $v^\mu$ denotes the four-velocity associated to the multiplet
$\hat J$; $\psi_\mu$ and $\eta$ are the spin 1 and spin 0 components
respectively; the radial quantum number has been omitted. Notice that
the multiplet $\hat J$ does not have indices related to light flavors.

The even parity $P-$wave quarkonium multiplet of states are described
by the matrix field~\cite{Casalbuoni:1992yd} ($\epsilon_{0123}=+1$):
\begin{equation}
J^{\mu}=\frac{1+\vsl}{2} \left(\chi_{2}^{\mu\alpha}\gamma_{\alpha}+
\frac{i}{\sqrt{2}} \epsilon^{\mu\alpha\beta\gamma}\chi_{1\gamma}v_{\alpha}\gamma_{\beta} +\frac{1}{\sqrt{3}}\chi_{0}(\gamma^\mu-v^\mu)+h^\mu\gamma_{5}\right)\frac{1-\vsl}{2}
\end{equation}
with $J_\mu v^\mu =0$. The  $\chi_2^{\mu \alpha}$, $\chi_1^\mu$, $\chi_0$ and
$h^\mu$ fields annihilate $\chi_{QJ}(nP)$ and $h_Q(nP)$ quarkonium states,
with $J^{PC}= 0^{++}, 1^{++}, 2^{++}$ and $1^{+-}$, respectively.  Note that
the spin two field is symmetric, traceless and orthogonal to $v^\mu$,
as $\chi_{1\mu}$ and $h_{\mu}$.  Under 
parity and charge conjugation symmetries, the matrix field $J^\mu$ transforms
as follows
\begin{eqnarray}
J^\mu(x^0,\vec{x}) &\stackrel{P}{\to} &\gamma^0 J_\mu(x^0,-\vec{x})
\gamma^0, \qquad v^\mu  \stackrel{P}{\to}    v_\mu \\
J^\mu &\stackrel{C}{\to} &c J^{\mu t} c
\end{eqnarray}
The hermitian
conjugate field $\bar{J_{\mu}}$ is defined as
\begin{equation}
\bar {J^{\mu}} = \gamma^0 J^{\mu\dagger} \gamma^0,
\end{equation}
and under heavy quark/antiquark rotations, we have
\begin{equation}
J_\mu  \to   S_QJ_\mu S^\dag_{\bar Q}, \qquad \bar J_\mu\to
S_{\bar Q }\bar J_\mu S_Q^\dag \label{eq:14}
\end{equation}
\subsection{$P^{(*)}\bar P^{(*)} \to P^{(*)}\bar P^{(*)}$ scattering}

At very low energies, the
interaction between a heavy and anti-heavy meson can be accurately
described just in terms of a contact-range potential. Pion
exchange  effects turn out to be
sub-leading~\cite{Nieves:2012tt,Valderrama:2012jv}. The LO Lagrangian respecting HQSS reads~\cite{AlFiky:2005jd}
\begin{eqnarray}
\label{eq:LaLO}
\mathcal{L}_{4H} & = & C_{A}\,\Tr\left[\bar{H}^{(Q)a}H_a^{(Q)} \gamma_{\mu}
\right] \Tr\left[{H}^{(\bar{Q})a} \bar{H}^{(\bar{Q})}_a \gamma^{\mu} \right]
\nonumber\\ &+&
C_{A}^{\tau}\,\Tr\left[\bar{H}^{(Q)a} \vec\tau_{\,\,.\,a}^{\,b}
{H}^{(Q)}_{b} \gamma_{\mu} \right] \Tr\left[{H}^{(\bar{Q})c}
\vec\tau_{\,\,.\,c}^{\,d}\bar{H}^{(\bar{Q})}_{d} \gamma^{\mu} \right]
\nonumber\\
&+& C_{B}\,\Tr\left[\bar{H}^{(Q)a}{H}_a^{(Q)} \gamma_{\mu}\gamma_5
\right] \Tr\left[{H}^{(\bar{Q})a} \bar{H}^{(\bar{Q})}_a \gamma^{\mu}\gamma_5
\right] \nonumber\\
&+&
C_{B}^{\tau}\,\Tr\left[\bar{H}^{(Q)a} \vec\tau_{\,\,.\,a}^{\,b}
{H}^{(Q)}_{b} \gamma_{\mu}\gamma_5 \right] \Tr\left[{H}^{(\bar{Q})c}
\vec\tau_{\,\,.\,c}^{\,d}\bar{H}^{(\bar{Q})}_{d} \gamma^{\mu}
\gamma_5\right] 
\end{eqnarray}
with $\vec\tau_{\,\,.\,a}^{\,b}$ the element $(a,b)$ [row,column] of the
Pauli matrices in isospin space, and $C_{A,B}^{(\tau)}$ light flavor independent
 LECs, which are also assumed to be heavy
 flavor independent and have dimensions of $E^{-2}$.  Note that in our normalization the heavy or anti-heavy meson
  fields, $H^{(Q)}$ or $H^{(\bar Q)}$, have dimensions of
  $E^{3/2}$ (see \cite{Manohar:2000dt} for details). This is because
  we use a non-relativistic normalization for the heavy mesons, which
  differs from the traditional relativistic one by a factor
  $\sqrt{M_H}$.
For later use, the four LECs that appear in Eq.~(\ref{eq:LaLO})  are rewritten into
$C_{0A}$, $C_{0B}$ and $C_{1A}$, $C_{1B}$ which stand for the
LECs in the isospin $I=0$ and $I=1$ sectors, respectively. The
relation between both sets reads
\begin{equation}
C_{0\phi} = C_{\phi} + 3 C_{\phi}^{\tau}, \qquad
C_{1\phi} = C_{\phi} - C_{\phi}^{\tau}, \qquad \text{for}~ \phi = A,B\ .
\end{equation}
\subsection{$Q\bar Q$ $n\, ^{2s+1}P_J$ quarkonium--$P^{(*)} \bar P^{(*)}$ transition}

There is only one HQSS consistent term describing the
LO interaction  of the $n\, ^{2s+1}P_J$  quarkonium states with the $P^{(*)} \bar
P^{(*)}-$pairs~\cite{Colangelo:2003sa}, 
\begin{equation}
\mathcal{L}_{HH
  Q\bar{Q}}=\frac{d}{2}\,\Tr[H^{a(\bar{Q})}\bar{J}_{\mu}H^{(Q)}_a\gamma^{\mu}]+\frac{d}{2}\,\Tr[\bar  H^{a(Q)} J_{\mu}\bar H^{(\bar Q)}_a\gamma^{\mu}] \label{eq:intlag}
\end{equation}
This expression accounts for the fact that the two heavy-light mesons
are coupled to the heavy-heavy state in $S-$wave, and therefore the
matrix elements do not depend on their relative momentum. Thanks to
HQSS, the same coupling controls the interaction of heavy-light mesons
both with the three $\chi$  states and also with the $h$ one. Another way to see that
the interaction term is unique is as follows. To describe the $S-$wave
molecular state, instead of using the basis in which the
meson-antimeson pair are coupled to a definite total spin state $\vert
j_{P^{(*)}}j_{\bar P^{(*)}}IJ\rangle$, with $I$ and $J$ the total
isospin and spin of the
system, one can choose a different basis in which the heavy and light
quarks are independently coupled to definite spins, and the
whole system is combined to make the definite spin of the whole
state. The elements of such basis are of the form $\vert (s_Q s_l) IJ
\rangle$, where $s_Q=0,1$ ($s_l=0,1$) is the spin of the heavy (light)
quark-antiquark pair, and $I$ the isospin of the configuration of the
light degrees of freedom. Only isoscalar $S-$wave molecular states
will be relevant for this discussion. The possible transitions between
isoscalar molecular and the quarkonia states can be described in terms
of the matrix elements of the form (for simplicity, we drop out the
isospin index)
\begin{equation}
\langle n ^{2s+1}l_{J'} \vert H^{QCD} \vert (s_Q s_l)
J\rangle = \delta_{J,J'}\delta_{s,s_Q}\langle nl \vert\vert  H^{QCD}
\vert\vert s_l\rangle
\end{equation}
where we have made use of rotational invariance and of HQSS, which
guaranties that the spin of the heavy-quark subsystem $s_Q$ is
conserved. Using charge conservation, it can also be shown that the
matrix element with $s_l=0$ is zero.  Indeed, charge conjugation in
the molecular states is given by $(-1)^{s_l+s_Q}$, which together with
the action of this symmetry, $(-1)^{1+s}$, on the $P-$wave quarkonium
states implies that only the $s_l=1$ matrix element is different from
zero\footnote{On can also argue that since $s_Q$ and $J$ are
  conserved, the remaining angular momentum, $\vec{J} - \vec{s}_Q$
  should also be conserved.  In the molecular state it corresponds to
  $s_l$ (since $L=0$ in the molecule), in the charmonium state
  $\vec{J} - \vec{s}_Q$ corresponds to $L=1$. Hence, conservation of
  $\vec{J} - \vec{s}_Q$ implies that only the $s_l=L=1$ matrix element
  is non-zero.}. 

The parameter $d$ in Eq.~(\ref{eq:intlag}) is an unknown LEC, with
dimensions of $E^{-1/2}$. It might depend on the radial quantum number
$n$, and it should be fitted to experimental data or be determined
otherwise. Moreover for a consistent treatment of mesons with two heavy
quarks, $1/m_Q$ corrections should also be included~\cite{Jenkins:1992nb}, breaking the
heavy quark symmetry. This leads to a possible dependence of the $d$ LEC  on the
heavy flavor configuration.  Other parameters which are
introduced into the model by the inclusion of the quarkonium degrees
of freedom are the masses of these new states.

Expressed in terms of the individual fields, the interaction
Lagrangian of Eq.~(\ref{eq:intlag}) reads:
\begin{eqnarray}
\mathcal{L}_{HH Q\bar{Q}}&=& -\sqrt{2}d\left[-\sqrt{2}\chi_1^{\dagger\eta}\left(P\bar
  P^*_\eta-P^*_\eta\bar P\right) -\sqrt{3}\chi_0^{\dagger}\left(P\bar P +
  \frac13P^*_\eta\bar P^{*\eta}\right)\right. \nonumber  \\ 
&+&\left. h^{\dagger\eta}\left(P\bar P^*_\eta+ P^*_\eta\bar P\right)
+i\epsilon_{\alpha\mu\rho\eta}v^{\alpha}h^{\dagger\mu}P^{*\rho}\bar
P^{*\eta}+2\chi_2^{\dagger\rho\eta} P^*_\rho \bar P^*_\eta\right] +
h.c. \label{eq:Lccbar}
\end{eqnarray}
where $P^{(*)}\bar P^{(*)}$ annihilates an isospin zero two meson 
state, normalized to 1. For instance in the case of charmed mesons,
the field combination would be 
\begin{equation}
\vert 00>=-\frac{1}{\sqrt{2}}\left(D^{0(*)}\bar
D^{0(*)}+D^{+(*)}D^{-(*)}\right). 
\end{equation}
Note that we use the
isospin convention $\bar u
= \vert 1/2,-1/2\rangle$ and  $\bar d
= -\vert 1/2,+1/2\rangle$, which induces $D^0=\vert
1/2,-1/2\rangle$ and $D^+=-\vert
1/2,+1/2\rangle$. 

\subsection{Charmonium radiative transitions}
\label{sec:char-rad}
As we shall see, the study of the $2P \to 1S, 2S$ charmonium radiative
transitions can help to constrain the mixing between the $D^{(*)}\bar
D^{(*)}$  and $2P$ charmonium degrees of freedom. We write the
Lagrangian for these radiative decays, within the dipolar
approximation, as follows~\cite{Casalbuoni:1992yd}
\begin{eqnarray}
\mathcal{L}_{\gamma} & = & \delta_n \,\Tr\left(\bar J_\mu(2P)\hat J(nS)
\right)v_{\nu}F^{\mu\nu} +h.c.\\
&=&\delta_nv^\nu F_{\mu\nu}\left\{2\eta_c^\dagger
h_{c}^\mu+2\chi_{2c}^{\mu\sigma}\psi^\dagger_\sigma(nS)+\frac{2\chi_{0c}}{\sqrt{3}}\psi^{\mu\dagger}(nS)
-i \epsilon^\mu_{.\sigma\alpha\beta}\psi^{\sigma}(nS)v^\alpha\chi_{c1}^\beta\right\}
\end{eqnarray}
where $n$ is the radial quantum number of the $0^{-+}$ and $1^{--}$
charmonium states described by the field $\hat J(nS)$, $F^{\mu\nu}$ is
the electromagnetic tensor and $ \delta_n $ is a dimensional parameter
($[E^{-1}]$), which also depends on the heavy flavor, at least
through the heavy quark electric charge.  The above Lagrangian conserves parity,
charge conjugation and it is invariant under HQSS transformations
since electric transitions do not change the quark spin. It is
straightforward to obtain for the E1 $\chi_{c1}(2P)\to \psi(nS)\gamma$
transition~\cite{Casalbuoni:1992yd}
\begin{equation}
\Gamma\left[\chi_{c1}(2P)\to \psi(nS)\gamma\right] =
\frac{\delta_n^2}{3\pi}E_\gamma^3\frac{M_{\psi(nS)}}{m_{\chi_{c1}}} \label{eq:rad1}
\end{equation}
where $E_\gamma$ is the photon energy. The comparison with the
expressions given in Ref.~\cite{Barnes:2005pb} leads to the
identification
\begin{equation}
\delta_n= \left(\frac{4\pi\alpha }{3}e^2_c\right)^{\frac12}\langle nS\vert r\vert 2P\rangle,\qquad
\langle nS\vert r\vert 2P\rangle= \int_0^{+\infty}dr r^2 R_{nS}(r) r R_{2P}(r)
\end{equation}
with $e_c=2/3$, the charm quark electric charge (in proton electric
charge units), and the normalization of the radial wave functions
given by
\begin{equation}
\int_0^{+\infty}dr r^2 R_{nL}(r) R_{n'L}(r) = \delta_{nn'}
\end{equation}

\section{Unitarized isoscalar amplitudes from HQSS LO potentials}
\label{sec:unitarized}
In this section, we  first give the isoscalar amplitudes obtained by
solving the LSE's in coupled channels using as kernels the potentials
deduced from the HQSS LO Lagrangians
discussed in the previous section. We particularize for the hidden charm
molecular and  $2P$ quarkonium states, though the extension to the
bottom case  is straightforward. 

For $D\bar D^*$, the $C$-parity states are $[D\bar D^*]_{\pm}=
(D\bar D^*\mp D^*\bar D)/\sqrt2$, 
and satisfy $C [D\bar D^*]_{\pm} = \pm
     [D\bar D^*]_{\pm}$. In our convention, the $C$-parity of these
     states is independent of the isospin and it is equal to $\pm
     1$. The relevant channels in the different $J^{PC}$ sectors are:
\begin{eqnarray}
\nonumber 
J^{PC}=0^{++}&:& \left\{ D\bar D, D^* \bar D^*, \chi_{c0}(2P)\right\} \\ \nonumber 
J^{PC} =1^{++}&:&\left\{\frac{1}{\sqrt2}\left(D \bar D^* -D^* \bar D\right), \chi_{c1}(2P)\right\} \\ \nonumber 
J^{PC}= 2^{++}&:& \left\{D^* \bar D^*, \chi_{c2}(2P)\right\} \\
J^{PC}=1^{+-}&:& \left\{\frac{1}{\sqrt2}\left(D^* \bar D+D \bar D^*\right),D^*
\bar D^*, h_{c}(2P) \right\}
\end{eqnarray}
\subsection{QM potentials}
From the Lagrangians of Eqs.~(\ref{eq:LaLO}) and (\ref{eq:Lccbar}), we
obtain Feynman amplitudes, $T^{\rm FT}$, which in turn are used to
define the non-relativistic Quantum Mechanics (QM) potentials, with the convention,
\begin{eqnarray}
V^{\rm QM}\left[D^{(*)}\bar D^{(*)}\to D^{(*)}\bar
  D^{(*)}\right] &=& \frac{T^{\rm FT}\left[D^{(*)}\bar D^{(*)}\to D^{(*)}\bar
  D^{(*)}\right]}{\sqrt{2M_{D^{(*)}}2M_{\bar
    D^{(*)}}2M_{D^{(*)}}2M_{\bar D^{(*)}}}}=-\frac{\mathcal{L}_{4H}}{4}  \\
V^{\rm QM}_{c\bar c}\left[\psi_{c\bar c}(2P)\to D^{(*)}\bar
  D^{(*)}\right] &=& \frac{T^{\rm FT}\left[\psi_{c\bar c}(2P)\to D^{(*)}\bar
  D^{(*)}\right]}{\sqrt{2\mbare 2M_{D^{(*)}}2M_{\bar D^{(*)}}}}=-\frac{\mathcal{L}_{HH
  Q\bar{Q}}}{2\sqrt{2}}
\end{eqnarray}
with $\psi_{c\bar c}$, the $\chi_{cJ}(2P)$ or $h_{c}(2P)$ charmonium
state, and $\mbare$ its common bare mass\footnote{Note that, here, by
  bare mass, we mean the mass of the charmonium states when the LEC
  $d$ is set to zero, $d=0$, and thus it is not a physical observable.
  Coupling to the $D^{(*)} \bar D^{(*)}$ meson pairs renormalizes this
  bare mass, as we will discuss below. Since, in the effective theory,
  the UV cut-off is finite, the difference between the bare and the
  physical charmonium masses is a finite renormalization. This shift
  depends on the UV regulator since the bare mass itself depends on
  the renormalization scheme.  The value of the bare mass, which is
  thus a free parameter, can either be indirectly fitted to
  experimental observations, or obtained from schemes that ignore the
  coupling of charmonium states to the mesons, such as some
  constituent quark models. In this latter case, the issue certainly
  would be to set the UV regulator to match the quark model and the
  EFT approaches.}.

The  isoscalar $\left[D^{(*)}\bar D^{(*)}\to D^{(*)}\bar
  D^{(*)}\right]$ potentials  have been obtained in
\cite{Nieves:2012tt}, 
\begin{eqnarray}
V^{\rm QM}(1^{++}) &=& C_{0A} + C_{0B} \label{eq:vqm1} \\
V^{\rm QM}(0^{++}) &=& 
\begin{pmatrix}
C_{0A} & \sqrt{3}\,C_{0B} \\
\sqrt{3}\,C_{0B} & C_{0A} - 2\,C_{0B}
\end{pmatrix} \, , \\
V^{\rm QM} (1^{+-}) &=&
\begin{pmatrix}
C_{0A} - C_{0B} & 2\,C_{0B} \\
2\,C_{0B} & C_{0A} - C_{0B}
\end{pmatrix} \, , \\
V^{\rm QM}(2^{++}) &=& C_{0A} + C_{0B} \, . \label{eq:vqm2}
\end{eqnarray}
Particle coupled-channel\footnote{We do not refer to
charge channels, but rather to the $P\bar P$ and $P^*\bar P^*$  or $P\bar P^*$ and $P^*\bar P^*$ mixings in the $0^{++}$ and $1^{+-}$
sectors, respectively.} effects turn out to be  sub-leading at
the charm and bottom scales~\cite{Nieves:2012tt,Valderrama:2012jv},
and  it was also the case for those due to pion exchanges. Hence, in
the phenomenological analysis carried out in Refs.~\cite{Guo:2013sya, HidalgoDuque:2012pq}, the off-diagonal elements of the
$0^{++}$ and $1^{+-}$ potentials were set to zero. However, in the
strict heavy quark limit, where pseudoscalar and vector heavy-light
mesons become degenerate, coupled-channel effects  need to be
considered. In that limit, and after diagonalizing the matrices, there
are appear two different eigenvalues $(C_{0a}-3C_{0b})$ and
$(C_{0a}+C_{0b})$, associated to the spin $s_l=0$ and $1$
configurations of the light
degrees of freedom, respectively. This gives rise to a large number of
degenerate molecular states in the heavy quark limit, as discussed in \cite{Hidalgo-Duque:2013pva,Baru:2016iwj}.

On the other hand, the $\left[\psi_{c\bar c}(2P)\to D^{(*)}\bar
  D^{(*)}\right]$ transition amplitudes are obtained from the Lagrangian of
Eq.~(\ref{eq:Lccbar}), 
\begin{eqnarray}
V^{\rm QM}_{c\bar c}(1^{++}) &=& d \qquad \qquad \qquad~ \chi_{c1}(2P) \to [D\bar
  D^*]_+ \label{eq:dinicial} \\ \nonumber \\
V^{\rm QM}_{c\bar c}(0^{++}) &=& - \frac{d}{2}
\begin{pmatrix}
\sqrt{3}  \\
1
\end{pmatrix} \qquad
\begin{pmatrix}
\chi_{c0}(2P) \to D \bar D\\
\chi_{c0}(2P) \to D^* \bar D^*
\end{pmatrix}  \\ \nonumber \\
V^{\rm QM}_{c\bar c} (1^{+-}) &=& - \frac{d}{\sqrt{2}}
\begin{pmatrix}
1 \\
1
\end{pmatrix} \qquad
\begin{pmatrix}
h_c(2P) \to  [D\bar
  D^*]_-\\
h_c(2P) \to D^* \bar D^*
\end{pmatrix}  \\ \nonumber \\
V^{\rm QM}_{c\bar c}(2^{++}) &=& d \qquad \qquad \qquad~ \chi_{c2}(2P)
\to D^*\bar D^* \label{eq:dfinal}
\end{eqnarray}
Due to the use of contact interactions, the LSE
shows an ill-defined UV behavior, and it requires a
regularization and renormalization procedure. We employ a standard
Gaussian regulator (see, {\it e.g.} ~\cite{Epelbaum:2008ga})
\begin{eqnarray}
\langle\vec{p}\,';\, D^{(*)}\bar D^{(*)}\,\vert V^{\rm QM}_\Lambda \vert \vec{p}\,;\, D^{(*)}\bar D^{(*)}\,\rangle & = & C_{0H}
~f_\Lambda(\vec{p}\,') f_\Lambda(\vec{p}\,) \\
\langle\vec{p}\,;\, D^{(*)}\bar D^{(*)}\,\vert V^{\rm QM}_{c\bar c; \Lambda} \vert \psi_{c\bar c}(2P)\rangle & \propto
&  d~ f_\Lambda(\vec{p}\,) \label{eq:defpot}
\end{eqnarray}
with $f_\Lambda(\vec{p}\,)=e^{-\vec{p}\,^{2}/\Lambda^{2}}$, 
$C_{0H}$ any of the combinations of isoscalar LECs that appear in
Eqs.~(\ref{eq:vqm1})--(\ref{eq:vqm2}), and the proportionality constants in
Eq.~(\ref{eq:defpot}) can be read off from
  Eqs.~(\ref{eq:dinicial})--(\ref{eq:dfinal}). We take cutoff values $\Lambda$
= 0.5--1 GeV \cite{Nieves:2012tt,HidalgoDuque:2012pq}, where the range
is chosen such that $\Lambda$ will be bigger than the wave number of the states,
but at the same time it will be small enough to preserve HQSS and
prevent that the theory might become sensitive to the specific details
of short-distance dynamics. The dependence of the results on the
cutoff, when it varies within this window, provides a rough estimate
of the expected size of sub-leading corrections. 

\subsection{Non-perturbative LSE re-summation}
\label{sec:LSE}
\begin{figure}[h]
\begin{center}
\includegraphics[width=1.0\textwidth]{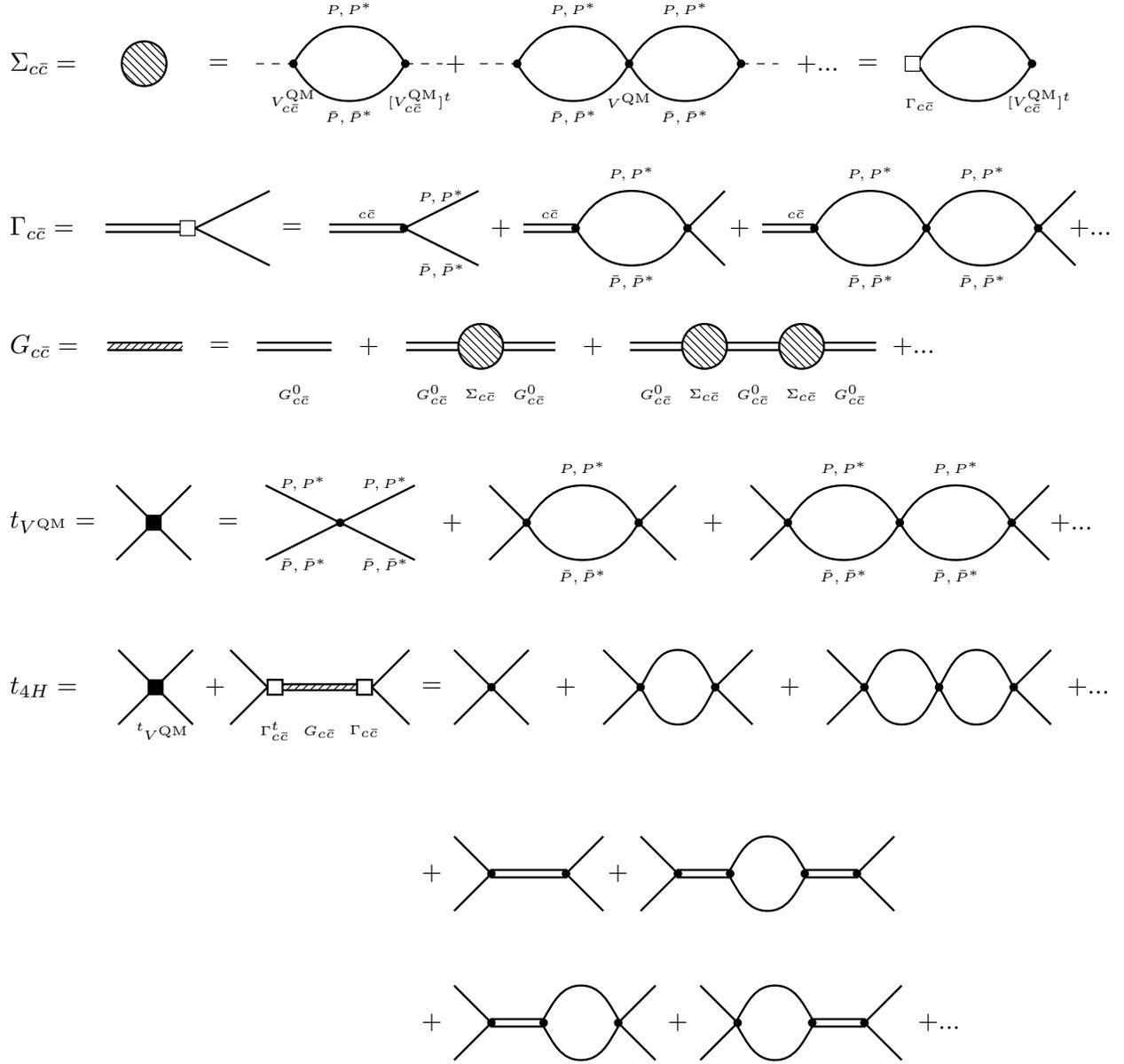}
\end{center}
\caption{Diagrammatic representation of different amplitudes: charmonium selfenergy
  ($\Sigma_{c\bar c}$), dressed charmonium 
  propagator ($G_{c\bar c}$) and  charmonium--$D^{(*)}\bar D^{(*)}$
  vertex function ($\Gamma_{c\bar c}$),  ``partial'' mesonic
  $t-$matrix ($t_{V^{\rm QM}}$), 
  and  full mesonic $t-$matrix ($t_{4H}$) defined in Eq.~(\ref{eq:defFullmesonT}).}\label{fig:1}
\end{figure}
The interplay of quark and meson degrees of freedom in a
near-threshold resonance was addressed in Ref.~\cite{Baru:2010ww}. We study physical states which are  mixture of a
$c\bar c$ bare state and some molecular components. Let us consider a
particular  $J^{PC}$
sector where there exist $n+1$ coupled channels, and assume that the
first $n$ channels are of molecular
type\footnote{We should nevertheless remind here once more that, 
  molecular coupled-channel effects should not be taken at LO for
  finite heavy quark masses, and that those effects appear at next-to-next
  leading order~\cite{Nieves:2012tt,Valderrama:2012jv}.}, while  the last one is $c\bar c$. The dynamics of such
system of energy $E$ is governed by a generalized $n+1$ dimension
$t-$matrix given by~\cite{Baru:2010ww} (diagrammatically, the most
relevant  elements are depicted in Fig.~\ref{fig:1}),  
\begin{equation}
\langle \vec{p}\,'\vert T(E) \vert \vec{p}\,\rangle=F_\Lambda(\vec{p}\,'\,)
\begin{pmatrix}
\left[t_{V^{\rm QM}}+\Gamma_{c\bar c} G_{c\bar c} \Gamma^{t}_{c\bar c}\right]_{n\times n} && \left[\dfrac{\Gamma_{c\bar c}}{1-G_{c\bar
    c}^0 \Sigma_{c\bar c}}\right]_{n\times 1} \\ \\
 \left[\dfrac{\Gamma^{t}_{c\bar c}}{1-G_{c\bar c }^0  \Sigma_{c\bar c
 }}\right]_{1\times n}  &&
\left[\dfrac{\Sigma_{c\bar c}}{1-G_{c\bar c }^0 \Sigma_{c\bar
      c}}\right]_{1\times 1}
\end{pmatrix} F_\Lambda(\vec{p}\,)
\label{eq:defT}
\end{equation}
where the Gaussian matrix of form factors reads\footnote{For on-shell
  mesons, the form-factor depends on the masses of the involved
  mesons, and hence on the meson channel. }
\begin{equation}
F_\Lambda(\vec{p}\,) = \begin{pmatrix}
{\rm Diag}\left[f_\Lambda(\vec{p}\,)\right]_{n\times n} & 0 \\ \\
 0 & 1
\end{pmatrix},
\end{equation}
On the other hand, the ``partial''  mesonic $t-$matrix\footnote{We call it ``partial'', because it does not
  incorporate $Q\bar Q $ effects on the
  meson-meson scattering.}, $t_{V^{\rm QM}}$ is
solution, once the Gaussian form-factor diagonal matrix
$f_\Lambda(\vec{p}\,)$ is also considered,  of a LSE with kernel
$V^{\rm QM}$, and it is given by
\begin{equation}
t_{V^{\rm QM}} = \left( 1-V^{\rm QM} G_{\rm QM}(E)\right)^{-1} V^{\rm QM}
\end{equation}
with $G_{\rm QM}(E)$, the diagonal meson loop function, conveniently regularized with
the Gaussian form-factor. For an arbitrary
energy $E$, its diagonal elements read~\cite{Albaladejo:2013aka}
\begin{align}
G_{\rm QM}(E) & = 
\int \frac{\text{d}^3 \vec{q}}{(2\pi)^3} \frac{e^{-2\vec{q}^{\,2}/\Lambda^2}}{E-M_1-M_2 - \vec{q}^{\,\,2}/2\mu + i0^+} \nonumber\\
& = -\frac{\mu\Lambda}{(2\pi)^{3/2}} + \frac{\mu k}{\pi^{3/2}}\phi\left(\sqrt{2}k/\Lambda\right)-i \frac{\mu k}{2\pi}e^{-2k^2/\Lambda^2}~,\label{eq:gmat_gr}
\end{align}
with $\mu^{-1}=M_1^{-1}+M_2^{-1}$, $k^2= 2\mu (E-M_1-M_2)$ and $\phi(x)$ the Dawson integral given by:
\begin{equation}
\phi(x) =  e^{-x^2}\int_{0}^{x} e^{y^2} \text{d}y~.
\end{equation}
Coming back to the different elements appearing in
Eq.~(\ref{eq:defT}), the non-relativistic bare $G_{c\bar c }^0$ and dressed $G_{c\bar c
}$ charmonium propagators are given by
\begin{equation}
G_{c\bar c }^0(E) = \frac{1}{E-\mbare }, \qquad
G_{c\bar c }(E) = \frac{1}{E-\mbare-\Sigma_{c\bar c}(E)}
\end{equation} 
where $\Sigma_{c\bar c}$ is the charmonium self energy induced by the meson
loops,
\begin{equation}
\Sigma_{c\bar c}(E) = \left[V_{c\bar c}^{\rm QM}\right]^t G_{\rm QM}(E)\Gamma_{c\bar c}(E)
\end{equation}
with the dressed vertex function, $\Gamma_{c\bar c}$, given by
\begin{equation}
\Gamma_{c\bar c}(E) = \left( 1-V^{\rm QM} G_{\rm QM}(E)\right)^{-1}V_{c\bar c}^{\rm QM}
\end{equation}

Two final remarks. First the $t-$matrix given in Eq.~(\ref{eq:defT})
can be also expressed as a solution of a LSE,
\begin{eqnarray}
\langle \vec{p}\,'\vert T(E) \vert \vec{p}\,\rangle&=&F_\Lambda(\vec{p}\,'\,)\, \left(\hat
V^{-1}- \hat G(E) \right)^{-1} F_\Lambda(\vec{p}\,) \label{eq:lsegral1}\\
\hat V &=& \begin{pmatrix}
V^{\rm QM} & V_{c\bar c}^{\rm QM} \\ \\
 \left[V_{c\bar c}^{\rm QM}\right]^t & 0 
\end{pmatrix}, \qquad \hat G(E) = \begin{pmatrix}
G_{\rm QM}(E) & 0 \\ \\
 0 & G_{c\bar c }^0(E)
\end{pmatrix}\label{eq:lsegral2}
\end{eqnarray}
and finally that, the full $(n\times n)-$mesonic $t-$matrix can be obtained as as
solution of a LSE equation with an energy dependent effective
potential $V_{\rm eff}(E)$~~\cite{Baru:2010ww},
\begin{eqnarray}
\langle \vec{p}\,'\vert t_{4H}(E)\vert \vec{p}\,\rangle&=&f_\Lambda(\vec{p}\,'\,)\,\left[t_{V^{\rm QM}}+\Gamma_{c\bar c} G_{c\bar c} \Gamma^{t}_{c\bar c}\right]f_\Lambda(\vec{p}\,)\nonumber\\
&=&  \left(f^{-1}_\Lambda(\vec{p}\,)\left[- G_{\rm QM}(E)+V^{-1}_{\rm eff}(E)\right]f^{-1}_\Lambda(\vec{p}\,'\,)\right)^{-1} \label{eq:defFullmesonT}\\
V_{\rm eff}(E) &=& V^{\rm QM}+ V_{c\bar c}^{\rm QM}G_{c\bar
    c}^0(E)\left[V_{c\bar c}^{\rm QM}\right]^t  \nonumber \\
&=&  V^{\rm QM}+ \frac{V_{c\bar c}^{\rm QM}\left[V_{c\bar c}^{\rm
      QM}\right]^t}{E-\mbare} 
\end{eqnarray}
In the strict heavy quark limit, where the full coupled-channel
effects should be considered, the effective matrix potential $V_{\rm eff}(E)$
gives rise to two different eigenvalues, $(C_{0a}-3C_{0b})$ and
$(C_{0a}+C_{0b}) +d^2/(E-\mbare)$. Thus, as compared to those
deduced from $V^{\rm QM}$, the interaction in the $s_l=0$ configuration has not been modified, while
the $s_l=1$ one is affected by the coupling to the quarkonium
states. The extra interaction becomes repulsive or attractive
depending on whether the energy $E$ is above or below the bare
charmonium mass, $\mbare$. Nevertheless we should stress, as
mentioned above, that in the present scheme  $\mbare$ is a free
parameter and it is not an observable, which gets dressed by the
$D^{(*)}-$meson loops and gives rise to the physical mass of the
charmonium states (see
for instance the discussion  below  Table~\ref{tab:dvsX} and
Fig.~\ref{fig:2} for the $1^{++}$ sector).

\section{Poles of the unitarized amplitudes and the compositeness condition  }
\label{sec:generalidades}
\subsection{Bound, resonant states and couplings}

The dynamically-generated meson states appear as poles of the
scattering amplitudes on the complex energy $E-$plane. The
poles of the scattering amplitude on the first Riemann sheet (FRS) that
appear on the real axis below threshold are interpreted as bound
states. The poles that are found on the second Riemann sheet (SRS) below the
real axis and above threshold are identified with resonances. The
mass and the width of the state can be found from the position of the
pole on the complex energy plane. Close to the pole, the scattering
amplitude behaves as
\begin{eqnarray}
T_{ij}&\sim&\frac{g_i g_j} {E-E_R}
\end{eqnarray}
The mass $M_R$ and width $\Gamma_R$ of the state result from $E_R= M_R-i\Gamma_R/2$, while $g_j$ (complex in general) is the
coupling of the state to the $j-$channel.

The meson loop function was given in Eq.~(\ref{eq:gmat_gr}). Note that
the wave number $k$ is a multivalued function of $E$, with a branch
point at threshold ($E=M_1+M_2$). The principal argument of
$(E-M_1-M_2)$ should be taken in the range $[0,2\pi[$. Note that this amounts to choosing the branch cut
      of the square root function defining $k$, to lie on the positive
      real line.  The function
    $k \phi(\sqrt2 k/\Lambda)$ does not present any discontinuity for
    real $E$ above threshold, and $G_{\rm QM}(E)$ becomes a
    multivalued function because of the $ i k $ term. Indeed, $G_{\rm
      QM}(E)$ has two Riemann sheets. In the first one, $0\leqslant
    {\rm Arg}(E-M_1-M_2)< 2\pi$, we find a discontinuity $G_{\rm
      QM}^I(E+ i\epsilon)-G_{\rm QM}^I(E-i \epsilon) = 2i\,{\rm Im}
    G_{\rm QM}^I(E+i\epsilon)$ for $E> (M_1+M_2)$. In the second
    Riemann sheet, $2\pi\leqslant {\rm Arg}(E-M_1-M_2)< 4\pi$, we
    trivially find $G_{\rm QM}^{II}(E- i\epsilon) = G_{\rm
      QM}^I(E+i\epsilon)$, for real energies and above threshold.

 \subsection{Components of the states and the compositeness condition} 
 
It is difficult to pin down the exact nature of a hadronic state since
wave functions are not observables themselves. The claims regarding
the largest Fock components in a wave function are often model
dependent. The compositeness condition, first proposed by Weinberg to
explain the deuteron as a neutron-proton bound
state~\cite{Weinberg:1962hj,Weinberg:1965zz}, has been advocated as a
model independent way to determine the relevance of hadron-hadron
components in a molecular state. However, this is strictly only valid
for bound states. For resonances, it involves complex numbers and,
therefore, a strict probabilistic interpretation is lost. The 
probabilistic interpretation of the compositeness condition has its
origin in the 
sum rule~\cite{Hyodo:2013nka, Sekihara:2014kya, Garcia-Recio:2015jsa}
\begin{equation}
-1= \sum_{ij} g_ig_j\left( \delta_{ij} \left[\frac{\partial
    G_i^{II}(E)}{\partial E}\right]_{E=E_R} + \left[ G_i^{II}(E)\frac{\partial
     V_{ij}(E)}{\partial E}G_j^{II}(E)\right]_{E=E_R}\right) \label{eq:sum-rule}
\end{equation}
which is satisfied by the residues of a pole, located in the fourth
quadrant of the SRS, of a $t-$matrix solution of a coupled--channel LSE,
\begin{equation}
T^{-1} = -G+V^{-1}
\end{equation}   
The above sum rule\footnote{We should note that
  Eq.~(\ref{eq:sum-rule}) is not the original Weinberg
  condition~\cite{Weinberg:1962hj,Weinberg:1965zz}, though it is
  undoubtedly inspired in the findings of those works.} is also satisfied in the case of bound states
(poles located in the real axis of the FRS below the lowest of the
thresholds) replacing $G^{II}\leftrightarrow G^I$. From
Eq.~(\ref{eq:sum-rule}), one might think that a possible definition of
the weight of a hadron-hadron component in a
composite particle could be
 \begin{equation}
  X_i= \mathrm{Re}\tilde{X}_i= \mathrm{Re}\left( -
  g_i^2\left[\frac{\partial G_i^{II}(E)}{\partial
      E}\right]_{E=E_R}\right) \label{eq:defxi}
 \end{equation}
As follows from the analysis in \cite{Aceti:2014ala} and \cite{Gamermann:2009uq}, for bound states, the
quantity $\tilde{X}_i$ is real and it is related to the probability of finding the state
in the channel $i$. For resonances, $\tilde{X}_i$ is still related to the
squared wave function of the channel $i$, in a phase
prescription that automatically renders the wave function
real for bound states, and so it can be used as a measure of
the weight of that meson-baryon channel in the composition
of the resonant state~\cite{Aceti:2014ala,Sekihara:2014kya}. 
The deviation of the sum of $X_i$ from unity is related to the energy
 dependence of the $S$-wave  potential,
 \begin{equation}
 \sum_i X_i=1-Z,  \label{eq:defXi}
 \end{equation}
 where 
 \begin{equation}
 Z={\rm Re}\tilde{Z}= {\rm Re}\left (-\sum_{ij} \left[g_i G_i^{II}(E)\frac{\partial
     V_{ij}(E)}{\partial
     E}G_j^{II}(E)g_j\right]_{E=E_R}\right) . \label{eq:defZ}
 \end{equation}
 Note that Eq.~(\ref{eq:sum-rule}) guaranties that the imaginary parts of
$\sum_i \tilde{X}_i$ and $\tilde{Z}$ must cancel.  The quantity $\tilde Z$, though complex in general, is defined even
for resonances, since it is related to the 
field renormalization constant~\cite{Hyodo:2011qc}
that is obtained by requiring that the residue of the renormalized two
point function will be one.
However its 
probabilistic interpretation is not straightforward.  Thus,
though $\tilde{X}_i$ can be interpreted as a probability of finding a
two-body component in a bound state, this interpretation, strictly
speaking, cannot be made in the case of a resonance. Nevertheless, because it represents the
contribution of the channel wave function to the total normalization,
the compositeness $\tilde{X}_i$ will have an important piece of
information on the structure of the resonance. Moreover,  in
Ref.~\cite{Guo:2015daa}, it was claimed 
that one can formulate a meaningful compositeness relation with only positive
coefficients thanks to a suitable transformation of the $S$
matrix. This in practice amounts to take the absolute value of
$\tilde{X}_i$ to quantify the probability of
finding a specific component in the wave function of a hadron.
Notice, however, 
that the recipe advocated in Ref.~\cite{Guo:2015daa} is not applicable
to all types of poles. In particular the arguments of this reference
exclude the case of virtual states or resonant signals which are an
admixture between a pole and an enhanced cusp effect by the pole
itself. More specifically, the probabilistic interpretation given in
\cite{Guo:2015daa} to
$|\tilde{X}_i|$ is only valid when ${\rm Re}(E_R) > M_{i,{\rm
    th}}$, with $M_{i,{\rm th}}$ the corresponding threshold of the
channel $i$.

For the present study, since the $V^{\rm QM}$ and $V_{c\bar c}$
potentials do not depend on the energy,  Eqs.~(\ref{eq:lsegral1}) and
Eq.~(\ref{eq:lsegral2}) should guaranty that the residues of the  poles of 
the on-shell $\langle \vec{p}\,'\vert T(E) \vert \vec{p}\,\rangle$ will fulfill 
\begin{equation}
-\sum_{i} g^2_i \left(\frac{\partial
    \left[\hat G_i(E)/F_{\Lambda\,i}^{2}\right]}{\partial E}\right)_{E=E_R} = 1 \label{eq:sume-rulebis}
\end{equation}
where the loop function should be computed in the FRS or SRS as
appropriate. Note that the above equation is not strictly correct, and
there exist minor corrections induced by the mild energy dependence
induced in the potentials inherited from the form-factor matrix $F_\Lambda(E)$.  We will make use of the
above sum-rule to address the molecular meson-meson  content of the
various poles obtained in the next subsection.

On the other hand, if we restrict ourselves to the full mesonic
$t-$matrix defined in Eq.~(\ref{eq:defFullmesonT}), we will face a
situation like that described in Eqs.~(\ref{eq:defXi}) and
(\ref{eq:defZ}). This is because, $t_{4H}$ is defined by means of an
energy--dependent effective potential result of integrating out the
quarkonium degrees of freedom. In this latter context, $X_i$ and $Z$
will be related to the weights of the two-body molecular and the
integrated out elementary (quarkonium) components, respectively.


\section{Quarkonium and the $1^{++}$ and $2^{++}$ meson molecules}
\label{sec:char-1++2++} 
HQSS predicts that in the heavy quark limit, the interaction in 
both the $1^{++}$ and $2^{++}$ sectors should be identical. Moreover, the
dynamics in these sectors
is governed by the $s_l=1$ configuration of the light degrees of
freedom, which is precisely that affected by the coupling between
quarkonium and meson-antimeson states. At the charm scale, we expect
some HQSS breaking effects due to the $D-D^*$, and the bare 
$\chi_{c1}(2P)-\chi_{c2}(2P)$ mass differences.

As mentioned in the introduction, assuming the $X(3872)$
to be a $D\bar D^*$ molecule, the existence of a $X_{2}$ [$J^{PC} =
  2^{++}$] $S$-wave $D^{*}\bar{D}^{*}$ bound state was predicted in
Refs.~\cite{Nieves:2012tt,HidalgoDuque:2012pq}, with
a binding energy similar to that of the $X(3872)$. The $X_2$ is not
affected by particle coupled channel effects and its mass only varies
mildly, by about 2--3 MeV, when corrections from the one pion exchange
potential are taken into account~\cite{Nieves:2012tt}. This prediction is subjected to some
uncertainties because of the approximate nature of HQSS. Hence,
the state might move slightly up above the $D^*\bar D^*$ threshold and
become virtual or might descend to a lower mass
region~\cite{Guo:2013sya}. Be as it may, one could be quite confident
about the existence of a molecular state with these quantum numbers
close to the $D^*\bar D^*$ threshold. However, the state has not been
observed yet.  

Within the EFT approach of
Refs.~\cite{Nieves:2012tt,HidalgoDuque:2012pq}, it is assumed that the
four-meson contact operator absorbs all the details of the short-range dynamics
present in the system, such as light vector meson exchanges between the
charmed mesons, or other Fock components in the $X(3872)$ and
$X_2(4012)$ wave functions. However, the effects due to the presence
of the $2P$ quarkonium states could be sizable, in particular in the
$1^{++}$ sector, because one expects the corresponding $c\bar c$ state to lie close to
the  $X(3872)$~\cite{Barnes:2005pb}. The experimental $\chi_{c2}(2P)$ mass,
$m_{\chi_{c2}}^{\rm exp}=3927.2
\pm 2.6$ MeV~\cite{Agashe:2014kda}, is significantly lower than the $D^*\bar D^*$ threshold,
and hence it looks reasonable to expect a limited  influence of the
charmonium level in  the dynamics of a loosely $2^{++}$ state
located in the vicinity of the $D^*\bar D^*$ threshold. However,
one should bear in mind that if the $\chi_{c1}(2P)$ is above the
$D\bar D^*$ threshold, but
relatively close to it, the presence of the charmonium state would provide an
effective attraction that will contribute to bind the $X(3872)$, but
it will not appear in the $2^{++}$ sector\footnote{Indeed, the
  $\chi_{c2}(2P)$  would provide an effective repulsion in this case, since it is
  placed below the $D^*\bar D^*$  threshold. Nevertheless, as commented
  before, the strength of such interaction would presumably be small 
  since the  $\chi_{c2}(2P)$ mass is significantly (90 MeV) lighter
  than the two body threshold.}. Because we are dealing with very weakly
bound states, it might well occur that these effects need to be
explicitly considered and they cannot be just accounted for in
short-distance LECs. This is what we want to qualitatively illustrate
in this section. To that end, and for simplicity, we work in the
isospin symmetric limit as done in
Refs.~\cite{Nieves:2012tt,Guo:2013sya} and in the  LQCD study of Ref.~\cite{Prelovsek:2013cra}, and use the averaged masses of
the heavy mesons, which are $M_D = 1867.24$~MeV, $M_{D^*} =
2008.63$~MeV, while we take the central value of the Particle Data
Group (PDG) averaged mass for the $X(3872)$,
$M_X=3871.69\pm0.17$~MeV~\cite{Agashe:2014kda}. We are aware of the
importance of the isospin breaking effects in the dynamics of this
resonance, specially in its strong decays, and we refer the reader to
Refs. ~\cite{HidalgoDuque:2012pq, Albaladejo:2015dsa} for a
comprehensive discussion. Taking into account such effects might obscure
the approach, which in this exploratory study needs to be qualitative,
because the existing uncertainties in the masses  of the bare
$\chi_{c1}(2P)$  and $\chi_{c2}(2P)$ states and in the value of the
LEC  that mixes meson-molecular and quarkonium components.

In the isoscalar $1^{++}$ and $2^{++}$ sectors (from now on, we will
be always referring to isoscalar sectors, but for the sake of brevity,
we will not explicitly mention it), the on-shell $t-$matrix of
Eq.~(\ref{eq:defT}) reads (we particularize it for the hidden charm
sectors, but its extension to the bottom ones is straightforward)
\begin{equation}
T(E)=
\dfrac{\Sigma_{c\bar c}}{1-G^0_{c\bar c}\Sigma_{c\bar c}} 
\begin{pmatrix}
f^2_\Lambda(E)\,\left[(d\, G_{\rm QM})^{-2}
  -\dfrac{1-G^0_{c\bar c}\Sigma_{c\bar c}}{G_{\rm QM}
    \Sigma_{c\bar c}}\right]\,\,\, & f_\Lambda(E)\,(d\, G_{\rm QM})^{-1}\\\\
f_\Lambda(E)\,(d\, G_{\rm QM})^{-1} & 1
\end{pmatrix} 
\label{eq:defT11}
\end{equation}
with  the on-shell form factor, $f_\Lambda(E)=
\exp\left\{-2\mu(E-M_1-M_2)/\Lambda^2\right\}$, and the quarkonium self energy given by
\begin{equation}
\Sigma_{c\bar c}(E) =\frac{d^{\,2}\, G_{\rm QM}(E)}{1-C_{0X}\,G_{\rm QM}(E)} 
\label{eq:eq60}
\end{equation}
where $C_{0X}= C_{0A}+C_{0B}$.
 The only differences between the $1^{++}$ and $2^{++}$ sectors are
 due to the meson and bare charmonium masses, which appear in the loop
 function, $G_{\rm QM}(E)$, $c\bar c$ bare propagator ($G^0_{c\bar c}$) and
 Gaussian form-factors. We use ($M_1= M_{D}$, $M_2= M_{D^*}$,
 $\mbareuno$) and ($M_1=M_2= M_{D^*}$, $\mbaredos$) for the $1^{++}$
 and $2^{++}$ sectors, respectively. As long as $d \ne 0$,
 poles\footnote{Note that when $d\to 0$, the $t-$matrix
   reduces to
\begin{equation}
 \lim_{d\to 0}T(E)=
\begin{pmatrix}
f^2_\Lambda\,\frac{C_{0X}}{1-C_{0X}G_{\rm QM}} &0 \\
0 & 0
\end{pmatrix} 
\nonumber
\label{eq:eq61}
\end{equation}
} of $T(E)$ correspond to zeros of the inverse of the dressed
 propagator
\begin{equation}
G_{c\bar c}(E_R)^{-1} = 0 \leftrightarrow 1-G^0_{c\bar c}(E_R)\Sigma_{c\bar
  c}(E_R) = 0, \qquad E_R = M_R-i\Gamma_R/2 \label{eq:pole-position}
\end{equation} 
in either the FRS (in that case $\Gamma_R\to 0^-$) or the SRS as appropriate. 
In the vicinity of the pole, we have in the corresponding Riemann sheet
\begin{equation}
\frac{\Sigma_{c\bar c}(E)}{1-G^0_{c\bar c}(E)\Sigma_{c\bar c}(E)} \sim
\frac{1}{E-E_R}\,\, \frac{\Sigma_{c\bar c}^2(E_R)}{1-\Sigma_{c\bar
    c}^\prime(E_R)},\qquad \Sigma_{c\bar
    c}^\prime(E_R) = \left.\frac{d\Sigma_{c\bar
    c}(E)}{dE}\right\vert_{E=E_R} \label{eq:vic}
\end{equation}
from what  follows that the couplings to the meson-antimeson
and bare charmonium states are,
\begin{eqnarray}
g_1^2 &=& \frac{\Sigma_{c\bar c}^2(E_R)}{1-\Sigma_{c\bar
    c}^\prime(E_R)}\,\frac{f_\Lambda^2}{d^{\,2}\left[G_{\rm
      QM}(E_R)\right]^2}= \frac{\Sigma_{c\bar
    c}^\prime(E_R)}{1-\Sigma_{c\bar
    c}^\prime(E_R)}\frac{f_\Lambda^2}{G^\prime_{\rm
      QM}(E_R)} \label{eq:defg1}\\
g_2^2 &=& \frac{\Sigma_{c\bar c}^2(E_R)}{1-\Sigma_{c\bar
    c}^\prime(E_R)} = \frac{(E_R-\mbare)^2}{1-\Sigma_{c\bar
    c}^\prime(E_R)} = -\frac{1}{1-\Sigma_{c\bar
    c}^\prime(E_R)}\,\frac{1}{G_{c\bar c}^{0\prime}(E_R)} \label{eq:rad2}
\end{eqnarray}
where
\begin{equation}
G_{\rm QM}^\prime(E_R) = \left.\frac{dG_{\rm
    QM}(E)}{dE}\right\vert_{E=E_R},  \qquad 
G_{c\bar c}^{0\prime}(E_R)= \left.\frac{dG^0_{c\bar c}(E)}{dE}\right\vert_{E=E_R} 
\end{equation}
On the other hand, Eq.~(\ref{eq:sume-rulebis}) is satisfied, and it
leads to
\begin{eqnarray}
g_1^2 \left(\frac{d\left[G_{\rm
      QM}(E)/f_\Lambda^2\right]}{dE}\right)_{E=E_R}+ g_2^2
\left(\frac{d\,G_{c\bar c}^0(E)}{dE}\right)_{E=E_R} &=& \frac{\Sigma_{c\bar
    c}^\prime(E_R)}{1-\Sigma_{c\bar
    c}^\prime(E_R)}- \frac{1}{1-\Sigma_{c\bar
    c}^\prime(E_R)}+ \cdots  \nonumber \\
&=& -1+\cdots \label{eq:compo}
\end{eqnarray}
where the corrections neglected above are of order ${\cal O}\left(\frac{f^\prime_\Lambda(E_R)/f_\Lambda(E_R)}{G^\prime_{\rm
      QM}(E_R)/G_{\rm QM}(E_R)}\right)$. These corrections, which for
$E_R=M_X$ are of the order of $5\%$, appear because the form-factor induces a mild energy dependence
in the $4H$ potential. As expected from the discussion of
Eq.~(\ref{eq:defZ}), we find 
\begin{eqnarray}
g_1^2 \left(\frac{d\left[V_{\rm
      eff}^{-1}(E)/f_\Lambda^2\right]}{dE}\right)_{E=E_R} = \frac{1}{1-\Sigma_{c\bar
    c}^\prime(E_R)}+ {\cal O}\left(\frac{f^\prime_\Lambda(E_R)/f_\Lambda(E_R)}{V^{-1\prime}_{\rm
      eff}(E_R)/V^{-1}_{\rm
      eff}(E_R)}\right)
\end{eqnarray}
Thus, in the $1^{++}$ and $2^{++}$ sectors we define the molecular ($\tilde X$) and
charmonium ($\tilde Z$)
probabilities, weights in general, of the pole placed at $E_R=
M_R-i\Gamma_R/2 $ as
\begin{equation}
\tilde X= -\frac{\Sigma_{c\bar
    c}^\prime(E_R)}{1-\Sigma_{c\bar
    c}^\prime(E_R)}, \qquad \tilde Z= \frac{1}{1-\Sigma_{c\bar
    c}^\prime(E_R)} \label{eq:defXZ}
\end{equation}
and $\Sigma_{c\bar c}^\prime(E_R)$ is given by
\begin{equation}
\Sigma_{c\bar c}^\prime(E_R)= \frac{G^\prime_{\rm
      QM}(E_R) (E_R-\mbare)^2}{d^{\,2} G^2_{\rm QM}(E_R)} \label{eq:defSigmaprima}
\end{equation}
from where, we trivially find that the resonance couples to the
charmonium state through the meson loops, 
\begin{equation}
g_2=d\,\frac{g_1}{f_\Lambda}\,  G_{\rm QM}(E_R) \label{eq:g2vsg1}
\end{equation}
Besides, we can fix $C_{0X}$ in the presence of the mixing LEC
$d$, by requiring the $X(3872)$ resonance to be a $1^{++}$
bound state located in the FRS below the 
$D\bar D^*$ threshold.  This leads to
\begin{equation}
C_{0X}= \frac{1}{G_{\rm QM}^I(M_X)}-\frac{d^{\,2}}{M_X-\mbareuno}\label{eq:defc0x}
\end{equation}
which leaves us with only three undetermined parameters, $d,\mbareuno$ and $\mbaredos$ for  the present
simultaneous analysis of the $1^{++}$ and $2^{++}$ sectors, including
mixing with charmonium states.

\subsection{Numerical results: $X(3872)$ and $\chi_{c1}(2P)$}
\label{sec:X3872}
One of the greatest uncertainties of the present approach is the mass
of the bare $\chi_{c1}(2P)$ state. This state has not been identified
yet, while most recent constituent quark models predict masses for the
$\chi_{c1}(2P)$ ranging from around 3947.4
MeV~\cite{Ortega:2010qq,Segovia:2013wma} to 3906
MeV~\cite{Ebert:2011jc}, including the value of 3925 MeV obtained in
the classic work of T. Barnes, S. Godfrey and
E. S. Swanson~\cite{Barnes:2005pb}.  However all these models
overestimate the measured mass of the $\chi_{c2}(2P)$  for
which these works report 3969, 3949 and 3975 MeV, 
respectively. (We expect small effects from the $D^*\bar D^*$
loops, as discussed above.)  In this exploratory study, we take
\begin{equation}
\mbareuno=3906~{\rm MeV}
\end{equation}
from Ref.~\cite{Ebert:2011jc}, since this work 
provides the closest prediction to the experimental  mass 
of the $\chi_{c2}(2P)$ state. Nevertheless, we should remind here that
the bare mass  depends on  the UV regulator, since it is not
a physical observable. Furthermore, and as we already mentioned, there
exists the major problem of choosing the appropriate scale  to match
the constituent quark model and the EFT. At this point, we have
adopted a pragmatic view, and thus predictions obtained with two different UV
cutoffs, spanning a physically motivated range of values, will be
presented. The expectation is that the UV regulator dependence will be
absorbed into the LECs and thus predictions for observables  at the
end could become at most mildly regulator dependent.
%
\begin{sidewaystable}
\centering
\begin{tabular}{c|ccc|cccc}
$d$ & $C_{0X}$ & $g^{X(3872)}_{D\bar
    D^*}$ & 
  $\tilde X_{X(3872)}$ & $\left(m_{\chi_{c1}},
    \Gamma_{\chi_{c1}}\right) $  & $g^{\chi_{c1}}_{D\bar
    D^*}$ &  $|\tilde X_{\chi_{c1}}|$ & $\tilde Z_{\chi_{c1}}$  \\
$[{\rm fm}^{1/2}]$ & $[{\rm fm}^2]$ & $[{\rm GeV}^{-1/2}]$ 
  & &  ${\rm [MeV]}$ & $[{\rm GeV}^{-1/2}]$ &  \\\hline
0. & $-0.789$  &0.90 & 1 & (3906.0,0) & 0. & 0. & 1.\\ 
0.05 & $-0.774$ & 0.89 &  0.98 & (3906.6, 1.9) &~~ $0.01+0.16\,i$~~ &~0.02~& $0.99+ 0.01\,i$
\\
0.1 & $-0.731$ & 0.87 &  0.92 & (3908.2, 7.9) &~~ $0.03+0.31\,i$~~ &~0.06~&
$0.96+0.05\,i$
\\
0.15 & $-0.659$ & 0.83 &  0.84 & (3910.5, 19.2) &~~ $0.07+0.44\,i$~~ &~0.14~& $0.92+0.11\,i$
\\
0.20 & $-0.559$ & 0.78 &  0.75 & (3912.4, 37.8) &~~ $0.14+0.56\,i$~~
&~0.23~& $0.87+0.19\,i$
\\
0.25 & $-0.429$ & 0.73 &  0.66 & (3912.0, 67.0) &~~ $0.24+0.65\,i$~~ &~0.36~& $0.82+0.31\,i$
\\
0.30 & $-0.271$ & 0.68 &  0.57 & (3903.9, 112.8) &~~ $0.38+0.73\,i$~~ &~0.55~& $0.77+0.50\,i$
\\
0.35 & $-0.084$ & 0.63 &  0.49 & (3864.5, 185.2) &~~ $0.63+0.85\,i$~~ &$>1$& $0.70+1.01\,i$
\\
$d^{\rm \,crit}$ & $\phantom{-}0.000$ & 0.61 &  0.47 & (3798.3, 209.4) &~~ $0.93+1.09\,i$~~ &$>1$& $0.53+2.12\,i$
\\
0.375 & $\phantom{-}0.020$ & 0.61 &  0.46 & (3754.4, 186.4) &~~
$1.21+1.37\,i$~~ &$>1$& $0.29+3.66\,i$
\\
0.3775 & $\phantom{-}0.031$ & 0.61 &  0.46 & (3701.6, 93.5) &~~ $2.19+2.39\,i$~~ &$>1$& $-0.44+12.27\,i$
\\
0.40 & $\phantom{-}0.132$ & 0.59 &  0.43 & (3827.1, 0) at SRS  &~~ $0.96$~~
&$\tilde X_{\chi_{c1}}<0$& $2.07$
\\
0.45 & $\phantom{-}0.376$ & 0.55 &  0.37 & (3850.9,0) at SRS  &~~ $0.63$~~ &$\tilde X_{\chi_{c1}}<0$& $1.52$
\\
0.5 & $\phantom{-}0.649$ & 0.51 &  0.32 & (3858.4,0) at SRS   &~~ $0.51$~~ &$\tilde X_{\chi_{c1}}<0$& $1.36$
\\
1.0 & $\phantom{-}4.963$ & 0.29 &  0.11 & (3869.7, 0) at SRS  &~~
$0.21$~~ &$\tilde X_{\chi_{c1}}<0$& $1.08$
\\
2.0 & $\phantom{-}22.217$ & 0.15 &  0.03 & (3871.3, 0) at SRS   &~~ $0.10$~~ &$\tilde X_{\chi_{c1}}<0$& $1.02$
\\
$d\gg d^{\rm \,crit}$ & $\sim \frac{d^{\,2}}{\mbareuno -M_X}$ & $ {\cal O}(1/d)$ & $
  {\cal O}(1/d^2)$   & ($M_X- {\cal
    O}(\frac{1}{d^2})$, 0) at SRS  &~~ ${\cal
    O}(1/d)$~~ &$\tilde X_{\chi_{c1}}=- {\cal
    O}(\frac{1}{d^2})$& $1 + {\cal O}(\frac{1}{d^2})$\\\hline
\end{tabular}
\caption{ Properties of the $1^{++}$
hidden charm poles as a function of $d$.  We solve  Eq.~(\ref{eq:pole-position})
with $\Lambda= 1.0$ GeV and for each value of $d$, $C_{0X}$ is determined from
 Eq.~(\ref{eq:defc0x}). The position of the $X(3872)$ is fixed at
  $M_X=3871.69$ MeV in the FRS. The $\chi_{c1}(2P)$ pole is located
  in the SRS. Finally, $d^{\rm \,crit}(\Lambda=1\,{\rm GeV})=
  \sqrt{\frac{M_X-\mbareuno}{G_{\rm QM}^I(M_X)}}=0.370$
  fm$^{1/2}$. }\label{tab:dvsX}
\end{sidewaystable}
%
\subsubsection{ Influence of the  $d$ LEC on the properties
  of the $1^{++}$ hidden charm poles}

In Table~\ref{tab:dvsX}, we show the properties of the poles found in
the $1^{++}$ hidden charm sector as a function of the mixing LEC
$d$. We solve Eq.~(\ref{eq:pole-position}) with an UV cutoff of
$\Lambda =1$ GeV, the qualitative pattern of the results is similar
for 500 MeV, though some quantitative differences appear, as can be
seen in Table~\ref{tab:500charm} of the Appendix. Note that
$C_{0X}=C_{0X}(\Lambda)$, and this dependence on the UV regulator
should cancel that of the meson loop propagator $G_{\rm QM}$
(Eq.~(\ref{eq:gmat_gr})), such that observables (resonances masses,
widths, meson-meson scattering lengths, etc..)
become independent of the UV regulator (see discussion in
\cite{Nieves:2012tt}), up to higher order terms. This is accomplished
by definition for the $X(3872)$ mass, but however there exist some
residual UV cutoff dependence in its coupling to the $D\bar D^*$
meson-pair (see Tables~\ref{tab:dvsX} and
\ref{tab:500charm}). The mixing parameter $d$ also depends
on $\Lambda$. Thus, when we say that both, 1 and 0.5 GeV, UV cutoffs
lead to a qualitative similar dependence on $d$, we mean this, not for
specific values of $d$, but for  results obtained for both
cutoffs with values of $d$ which give rise to similar  meson-molecular
probabilities for the $X(3872)$
resonance ($\tilde X_{X(3872)}$).

In principle, we expect to find two poles\footnote{In the SRS, the
  poles appear as conjugate pairs \cite{Hanhart:2014ssa} if they are not on the real axis. We count
  these as single poles since they correspond to the same resonance.},
which will be identified as the $X(3872)$ and the physical
$\chi_{c1}(2P)$ states. Because of the election of $C_{0X}$ in
Eq.~(\ref{eq:defc0x}), the position of the $X(3872)$ is fixed at
$M_X=3871.69$ MeV, while its molecular probability ($\tilde
X_{X(3872)}$) and the $D\bar D^*$ coupling decrease with $d$. This is
because $C_{0X}$ absorbs all dependence on $d$, since $G_{\rm
  QM}^I(M_X)$ accounts only for the unitary logarithms and it is
independent of this LEC within the UV-cutoff scheme adopted here,
which guaranties that $\Sigma_{c\bar c}^\prime(E_R)$ in
Eq.~(\ref{eq:defSigmaprima}) scales as $1/d^2$.

\begin{figure}[h]
\begin{center}
\makebox[0pt]{\includegraphics[width=0.4\textwidth]{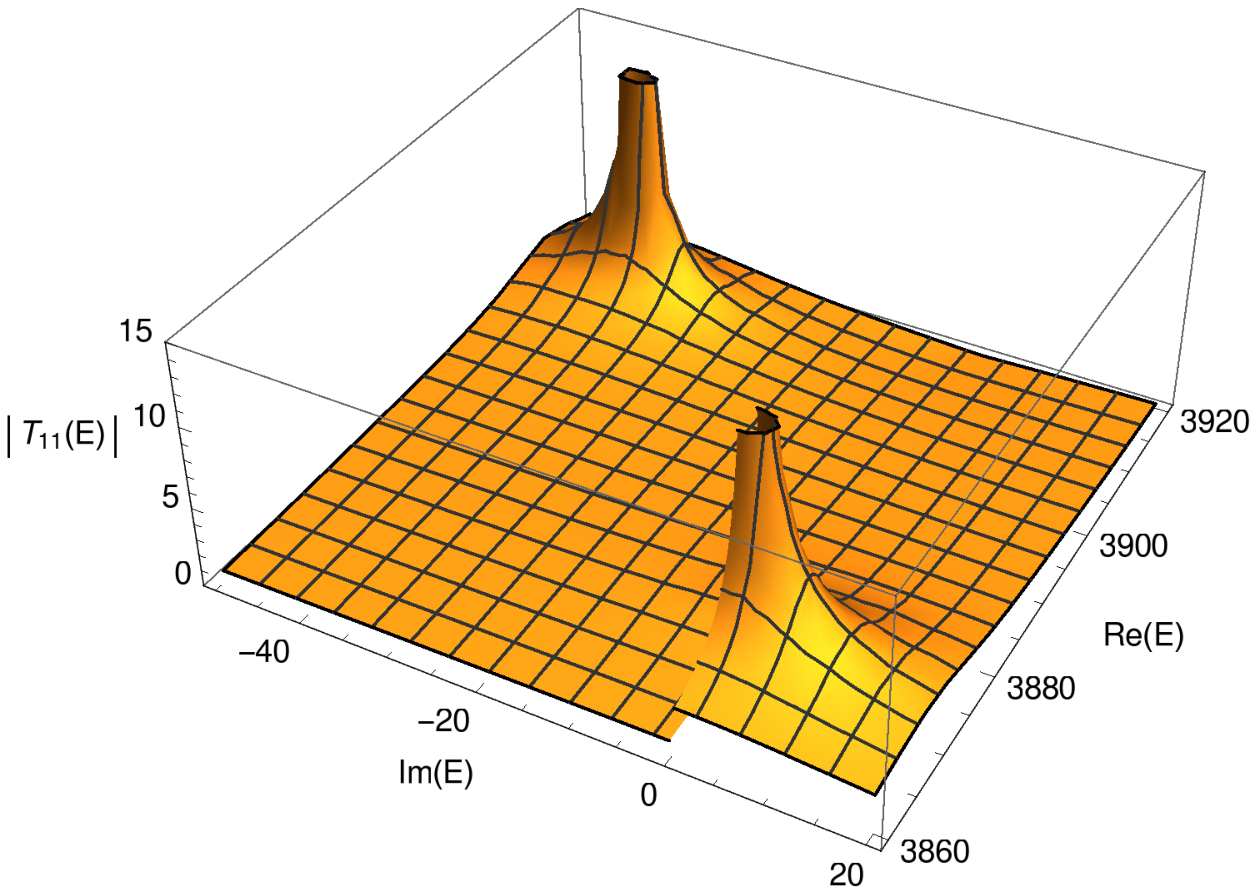}\hspace{0.75cm}\includegraphics[width=0.4\textwidth]{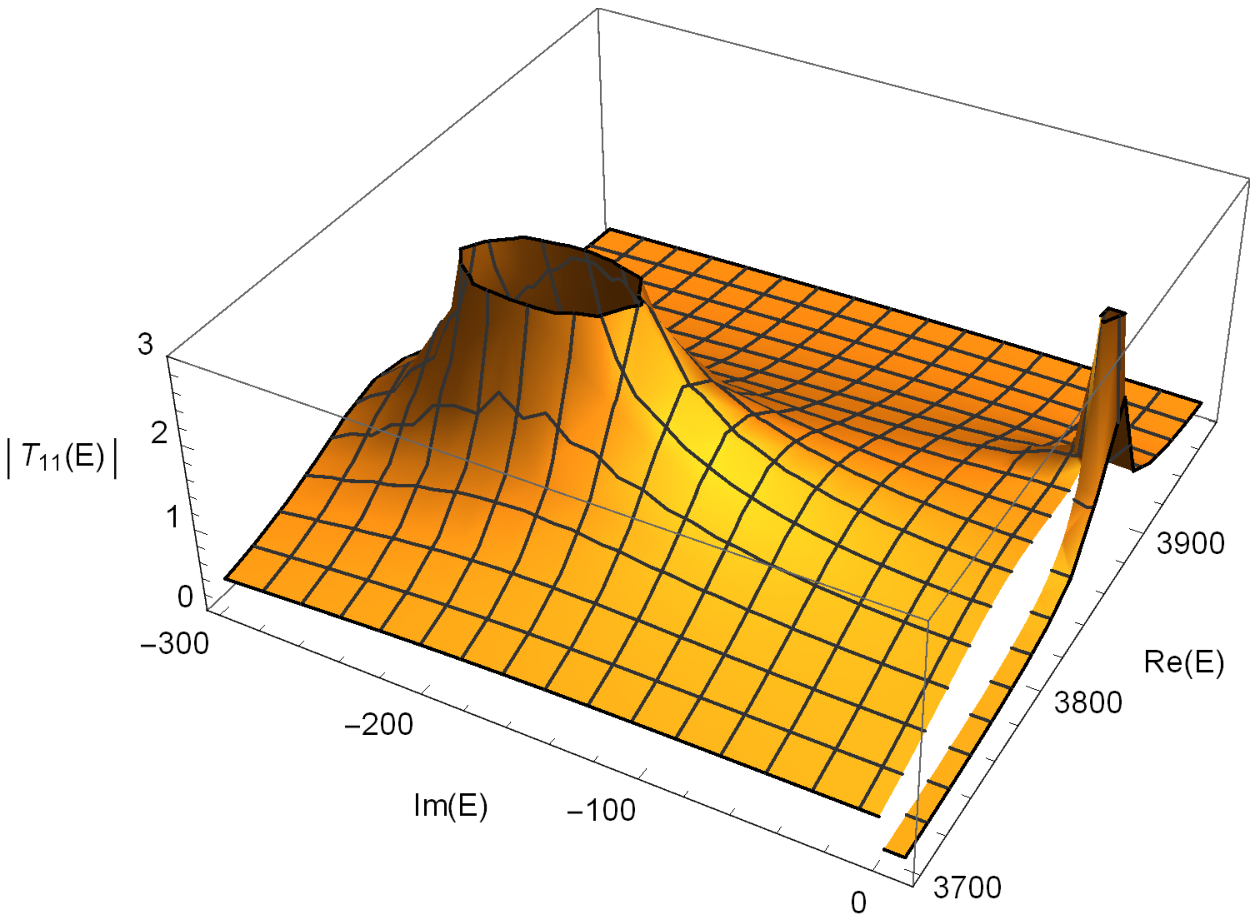}}\\\vspace{0.5cm}
\makebox[0pt]{\includegraphics[width=0.4\textwidth]{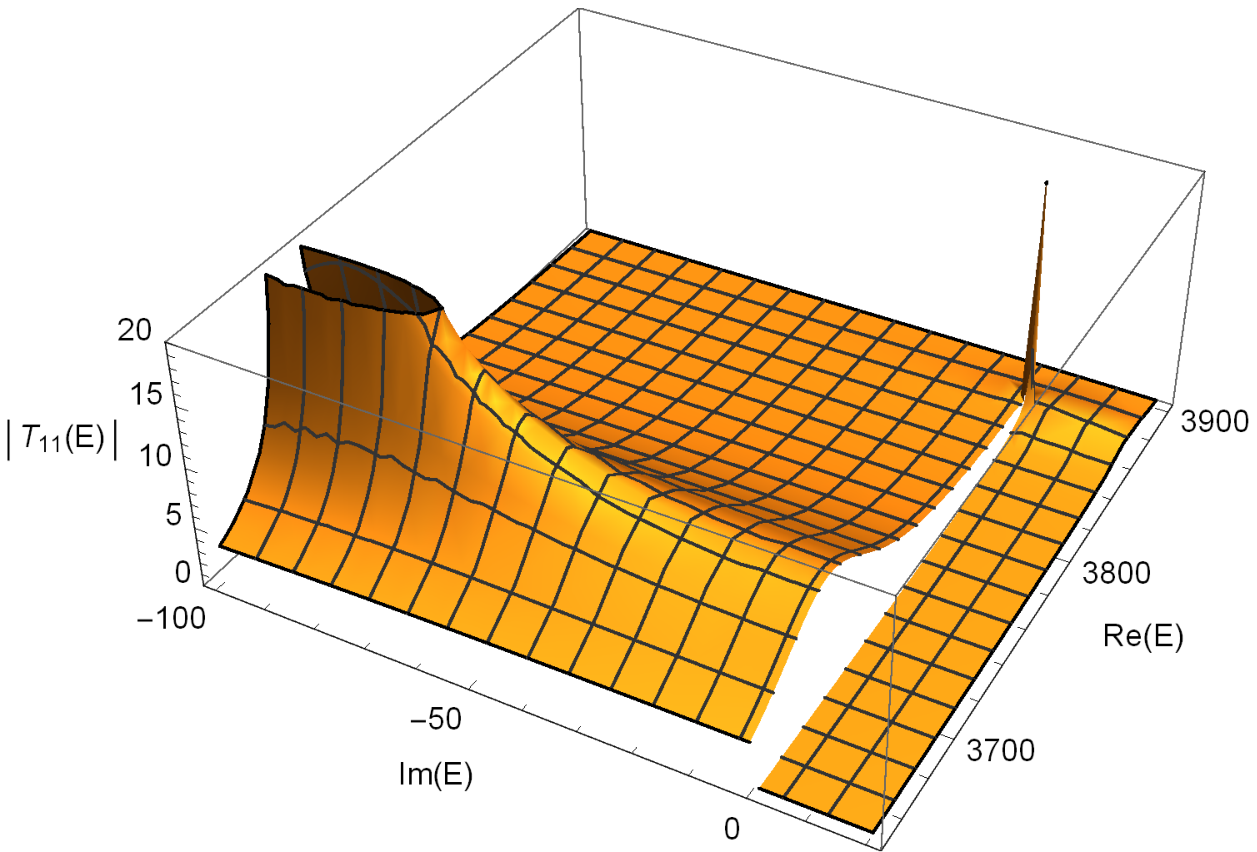}\hspace{0.75cm}\includegraphics[width=0.4\textwidth]{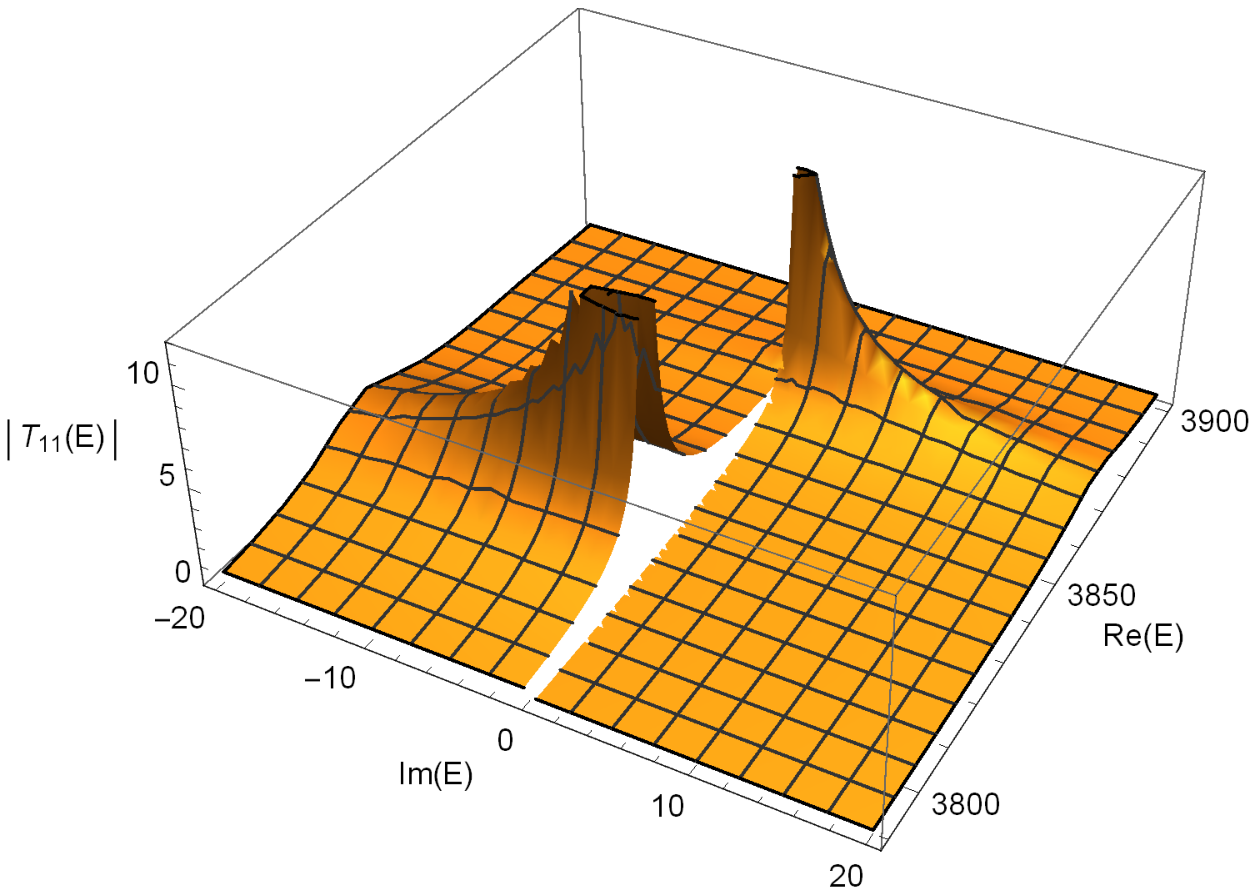}}\\\vspace{0.5cm}
\makebox[0pt]{\includegraphics[width=0.5\textwidth]{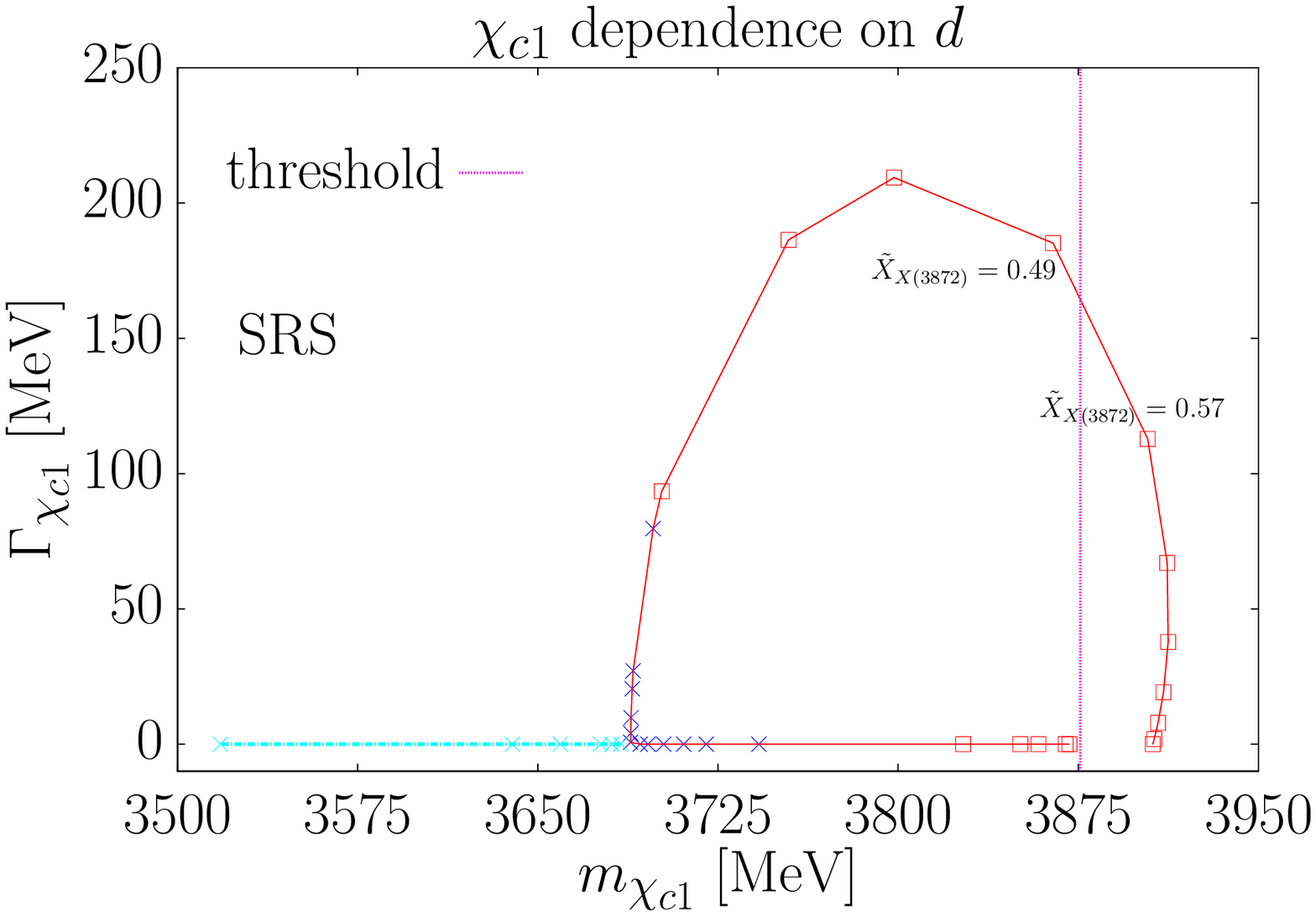}}
\end{center}
\caption{Hidden charm $J^{PC}=1^{++}$ sector. Top and middle panels:
  FRS (${\rm Im}(E)> 0$) and SRS (${\rm Im}(E)< 0$) of $|T_{11}(E)|$
  [fm$^2$] (Eq.~(\ref{eq:defT11})) as a function of the complex energy
  $E$ [MeV], for $d=0.20, d^{\rm \,crit}, 0.3775$ and $0.40$
  fm$^{1/2}$. Note that, since the $T-$matrix  is shown for only half
  of the SRS (and also the FRS), the pole in the SRS conjugate to the
  pole shown in the figures is not visible. Bottom panel: dependence of
  the $\chi_{c1}(2P)$ mass and width on $d$. Squares stand for the
  results of Table~\ref{tab:dvsX} at different values of $d$, while
  the crosses illustrate the highly non-linear behavior that appears
  when $d$ takes values in the interval [0.3776,\, 0.3785]
  fm$^{1/2}$. In this latter case, when the pole reaches the real
  axis, we find two poles, which start separating from each other and
  move apart from the ``meeting point'' (intersection with the real
  axis). Note that, no information about the pole that departs from
  threshold (cyan crosses) is given in Table~\ref{tab:dvsX}.  The curve is smooth except at the point where the pole
  hits the real axis on the SRS, however it looks like a broken line
  because the points are connected by straight segments.  All
  calculations have been carried out with an UV cutoff $\Lambda= 1$
  GeV. }\label{fig:2}
\end{figure}

On the other hand, the mass and the width of the $\chi_{c1}(2P)$
dressed state strongly depend on $d$. For moderate values of this LEC,
up to $\tilde X_{X(3872)} > 0.57$, the pole stays in the SRS above
threshold with its width increasing rapidly (f.i. top left panel of
Fig.~\ref{fig:2}). There is a point in the vicinity of $d^{\rm
  \,crit}$, value of the LEC for which $C_{0X}$ is zero, where the
$\chi_{c1}(2P)$ pole becomes below threshold and quite wide. Since
SRS and FRS are disconnected below threshold, such virtual state
becomes irrelevant (f.i. top right and middle left panels of
Fig.~\ref{fig:2}). When $C_{0X}=0$, the pole position equation reduces
to
\begin{equation}
E_R = \mbareuno +\left( M_X-\mbareuno \right) \frac{G_{\rm QM}(E_R)}{G^I_{\rm QM}(M_X)},
\qquad  E_R=M_R-i\Gamma_R/2
\end{equation}
which, besides $E_R=M_X$ in the FRS, has solutions in the SRS, but below
threshold. When $C_{0X}$ becomes positive (repulsive), the pole moves fast to the real axis
because there exist solutions only when $M_R < \mbareuno$ and
\begin{equation}
\frac{C_{0X}}{d^2} \left( \left( M_R-\mbareuno\right)^2 + \frac{\Gamma_R^2}{4}\right)\le |M_R-\mbareuno| 
\end{equation}
as deduced from the imaginary part of Eq.~(\ref{eq:pole-position}),
taking into account that Re$\left(G_{\rm QM}^{II}\right)<0$ in this
region. The intersection with the SRS real axis occurs for $d\sim
0.377823$ fm$^{1/2}$ that gives rise to a pole at $E_{R}= M_{0R}-i0$,
with $M_{0R} \sim 3688.67$ MeV. It turns out that in this intersection
$\Sigma_{c\bar c}^\prime(M_{0R}-i 0)= 1$ leading to singularities in
$\tilde X_{\chi_{c1}}$ and $\tilde Z_{\chi_{c1}}$, and provoking that
not only the inverse of the dressed propagator has a zero in this
intersection $\left[G_{c\bar c}(M_{0R}-i 0)^{-1} = G^0_{c\bar
    c}(M_{0R})^{-1}-\Sigma_{c\bar c}(M_{0R}-i 0) = 0\right] $, but
also its first derivative, ie. $dG^{-1}_{c\bar c}(E)/dE|_{E=M_{0R}-i
  0}=0$. Indeed, it is a double pole (see Eq.~(\ref{eq:vic})) since,
as mentioned above, the poles appear as conjugate pairs, which
obviously coincide in the real axis producing a kink.  Once the poles
collide on the real axis, they do not need to remain as a conjugate
pair. Indeed, as one pole approaches the threshold, with
$\Sigma_{c\bar c}^\prime$ decreasing and departing from 1, a second
pole moves away from the threshold, with now $\Sigma_{c\bar c}^\prime$
taking values above 1. (This behavior coincides with that discussed in
Fig.3 of Ref.~\cite{Hanhart:2014ssa}).  When $d~\sim 0.37854$
fm$^{1/2}$, this second pole leaves the real axis forming another
conjugate pair, with a mass of around 3470 MeV quite far from
threshold. The trajectories of this new conjugate pair as $d$
increases are either below threshold, or above threshold, but in this
latter case very deep in the complex plane\footnote{Actually, this
  latter part of the trajectory  could even be just an artifact of the model.} (widths of around 1
GeV). Hence, these poles will not have any observable consequences,
and for simplicity, we will  simply ignore them, and we have neither
included their  details in Table~\ref{tab:dvsX}. Actually in what
follows, we will always refer to the pole that moves along the real axis towards
threshold. Once, this pole has reached the SRS real axis (f.i. middle
right plot of Fig.~\ref{fig:2}), its position, $M_R$, is solution
(below threshold) of
\begin{equation}
(M_R-\mbareuno)\,\left(\frac{1}{G_{\rm
    QM}^{II}(M_R)}-\frac{1}{G_{\rm QM}^I(M_X)}\right) = d^2\,\left( 1-
\frac{{M_R-\mbareuno}}{{M_X-\mbareuno}}\right) \label{eq:lim}
\end{equation}
which differs from $M_X$ because $G_{\rm QM}^I(M_X)\ne G_{\rm
  QM}^{II}(M_X)$.  This non-trivial
$d-$behaviour is illustrated in the bottom panel of
Fig.~\ref{fig:2}. Note also $\Sigma_{c\bar c}^\prime(M_X+i0)$ and
$\Sigma_{c\bar c}^\prime(M_R-i0)$ have different signs. Since the pole
now becomes quite close to the threshold, where both SRS and FRS are
connected, it might have visible effects in scattering observables,
though its molecular content and the square of the coupling to the
$D\bar D^*$ scale as ${\cal O}(1/d^2)$. The same occurs for the
$X(3872)$, which in the $d\gg d^{\rm \,crit}$ limit appears to be a
charmonium state, mirror in the FRS of the pole found in the SRS. This
behaviour is in good agreement with the findings of
Ref.~\cite{Baru:2003qq} obtained using quite general arguments (see
discussion after Eq.~(22) of this latter reference). 
\newpage

The fact that in the limit $d \gg d^{crit}$, both poles become
dominantly charmonium can also be understood as follows. In order to
keep the position of the pole corresponding to the $X(3872)$ fixed, as
$d$ increases, $C_0$ should also increase and take large positive
values\footnote{Note that, this variation of $C_0$ depends on the
  procedure  used to renormalize the amplitudes. Since the position of
  the pole corresponding to the
  $X(3872)$ is fixed, from Eq. (\ref{eq:eq61}), one deduces  that the
  value of $\Sigma(M_X)$ is also fixed. In the regularization scheme
  used in this work, $G_{\rm QM}$ is independent of $d$, and hence from
  Eq. (\ref{eq:eq60}), it is clear that for large values of $d$, $C_0
  \propto d^2$. Furthermore, since $G_{\rm QM}$ is independent of $d$,
  Eq. (\ref{eq:defSigmaprima}) dictates that $\Sigma'_{c\bar c}
  \propto 1/d^2$ and hence $\tilde Z \simeq 1$ and $\tilde X
  \simeq 0$, i.e. we have a dominantly charmonium state.  An
  alternative scheme would be to keep $C_0$ fixed, but change the
  regularization of the loop function to keep the position of
  $X(3872)$ fixed.  In such  scheme, $G_{\rm QM} \propto \frac{1}{d^2}$,
  as can be seen from Eq. (\ref{eq:eq60}). This would be accomplished
  by means of an appropriate subtraction in the loop function, which
  would effectively account for some higher order terms in the interaction. In this scheme,
  $\Sigma'_{c\bar c} \propto d^2$, and hence $\tilde Z \simeq 0$ and
  $\tilde X \simeq 1$. However, one should bear in mind that the connection between the
  factors $\tilde X$ and $\tilde Z$ and the weights of the
  wave-functions of the various components in the state~\cite{Aceti:2014ala,Gamermann:2009uq} is inspired in
  the findings of the works of
  Ref.~\cite{Weinberg:1962hj,Weinberg:1965zz} by Weinberg. These
  latter results were found within
  non-relativistic quantum mechanics and for weakly bound
  states. Undoubtedly, the connection is clearer when an  UV cutoff is used to suppress the contribution of momenta much
  higher than the wave-number associated to the bound state.}.  These large positive values create
a strong repulsive contact force between the $D$ and $D^{*}$
mesons. This strong repulsive force, suppresses the contribution of
the molecular component in the states.

Results for larger (smaller\footnote{Note that
  Ref.~\cite{Ebert:2011jc} provides one of the smallest $\chi_{c1}(2P)$ bare
  masses among all recent predictions available in the literature.}) values of $\mbareuno$ are qualitatively
similar, though larger (smaller) $d$ values are needed to reach the
same amount of  charmonium component ($\tilde Z_{X(3872)}=1-\tilde X_{X(3872)}$) in the $X(3872)$. 
%

\subsubsection{Radiative decays of the $X(3872)$ and its charmonium content} 
\label{sec:rad-decays}
Using vector meson dominance and assuming that the $X(3872)$
is a hadronic molecule, with the dominant component $D^0¯D^{*0}$ plus a
small admixture of the $\rho J\psi$ and $\omega J/\psi$, the ratio of the $\X$
branching fractions into $\psi(2S)\gamma$ and $J/\psi\gamma$ 
 was calculated in~\cite{Swanson:2004pp} to be about
$4\times 10^{-3}$, which strongly differs from the
experimental value quoted in Eq.~\eqref{eq:Rexprad}.  In sharp
contrast, quark model
calculations, assuming a $c\bar c$ $2\,^3P_1$ nature for the $X(3872)$, predict a
wide\footnote{The results for $\X \to J/\psi \gamma$ are particularly
  sensitive to quark model details (see for instance Table 2 of
  Ref.~\cite{Swanson:2004pp}).} range for this ratio, where the
experimental ratio can be easily accommodated. 

\newpage

As mentioned in the Introduction, the study
of Ref.~\cite{Guo:2014taa} suggests that for  radiative decays
of the  $X(3872)$, short-range contributions are of similar importance
as their long-range counter parts, and that the measured value for $R_{\psi\gamma}$ is not in conflict with a
predominantly molecular nature of the $X(3872)$. Triangular
$D D^{(*)}\bar D^{(*)}$ and $D\bar D^*$ loop contributions to these
radiative decays  were computed in \cite{Guo:2014taa} (Figs. 1(a)-1(e)
of that reference), 
using dimensional regularization with the $\overline{{\rm MS}}$ subtraction
scheme at various scales $\mu=M_X/2,M_X,2M_X$. 
The results of Table 2 of Ref.~\cite{Guo:2014taa} can be summarized as follows
\begin{eqnarray}
\Gamma^{\rm loops}(\X\to J/\psi \gamma) &=& \left(9.7  + 19.9
\log\frac{2\mu}{M_X}\right)(r_xr_g)^2\,[{\rm keV}]\label{eq:XJpsi1}\\
\Gamma^{\rm loops}(\X\to \psi(2S) \gamma) &=& \left(3.8  + 1.6
\log\frac{2\mu}{M_X}\right)(r_xr'_g)^2\,[{\rm keV}] \label{eq:XJpsi2}
\end{eqnarray}
where we have adjusted the two lower values, $\mu=M_X/2$ and
$\mu=M_X$, given in the table for each decay mode. The interpolating
function works quite well in the case of the $\psi(2S)\gamma$ mode,
while it underestimates by around 15\% the width obtained in
\cite{Guo:2014taa} for the $J/\psi\gamma$ decay at $\mu= 2M_X$. In the
above expressions, $r_x= g_{XD\bar D^*}/(0.97$ GeV$^{-1/2})$, $r_g= g/(2$
GeV$^{-3/2})$ and $r'_g= g'/(2$ GeV$^{-3/2})$, with $g$ and $g'$, the
spin-symmetric $J/\psi D^{(*)} \bar D^{(*)}$ and $\psi(2S) D^{(*)} \bar D^{(*)}$
coupling constants (see Eqs.~(10)--(12) of
Ref.~\cite{Guo:2014taa}). Here we find in Tables~\ref{tab:dvsX} and \ref{tab:500charm}, 
 $g_{XD\bar D^*}=0.90$ GeV$^{-1/2}$ and 1.05 GeV$^{-1/2}$ for
$\Lambda=1$ and 0.5 GeV, respectively. Hence, the estimate taken in
Ref.~\cite{Guo:2014taa} is reasonable for the qualitative purposes of
the current work.  The  $J/\psi$ and
$\psi (2S)$ coupling constants to the charmed mesons cannot be measured directly and are
badly known. The value of 2 GeV$^{-3/2}$ for $g$ was taken in
\cite{Guo:2014taa} from the model estimates of Refs.~\cite{Guo:2010ak,
Colangelo:2003sa}. The estimate of $g'$ used to produce the central
values of Table 2 in Ref.~\cite{Guo:2014taa} is just an educated guess, though values
of $g'/g\sim 1.67$ are justified in the analysis of Ref.~\cite{Dong:2009uf}.

The charmed meson loop contributions to the $\Gamma(\X\to \psi(nS)
\gamma)$ decay show an important scale dependence, in particular in
the $J/\psi$ mode. Indeed, the ratio of the $\X$ branching fractions
into $\psi(2S)\gamma$ and $J/\psi\gamma$ calculated in
Ref.~\cite{Guo:2014taa} lies in the interval (0.14--0.39)$(g'/g)^2$,
being the $\psi(2S)$ channel suppressed, although a lot less than
claimed in Ref.~\cite{Swanson:2004pp}. (Note that values of $g'/g\sim
2$ would bring the ratio to be of order one in this purely molecular
picture). This supports the claim made in \cite{Guo:2014taa}
that for the radiative decays of the $X(3872)$ short-range
contributions are important.

\begin{figure}
\begin{center}
\includegraphics[width=0.8\textwidth]{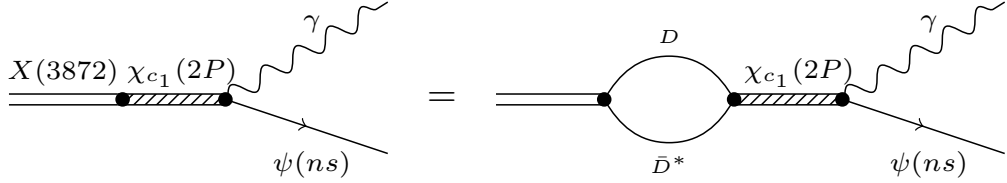}
\caption{Decay mechanism for the transition $\X\to \psi(nS)$ through
  an intermediate charmonium $\chi_{c1}(2P)$ state. The identity
  between the two diagrams follows from the relation between couplings
  in  Eq.~(\ref{eq:g2vsg1}).  }\label{fig:rad}
\end{center}
\end{figure}

 Since any physical amplitude should be independent of the scale, the
 dependence displayed in Eqs.~(\ref{eq:XJpsi1}) and (\ref{eq:XJpsi2})
 should be compensated by a corresponding variation in the
 counter-term contribution depicted in diagram 1(f) of
 Ref.~\cite{Guo:2014taa}. Since the counter-terms parametrize
 short-range physics they may be modeled by a charm quark loop.
 Hence, we could estimate the size of the counter-term by employing
 the model presented in this work and depicted in Fig.~\ref{fig:rad}.
From Eqs.~(\ref{eq:rad1}) and  (\ref{eq:rad2}), one trivially finds
\begin{equation}
\Gamma\left[\X\to \psi(nS)\gamma\right] =
\frac{(M_X-\mbareuno)^2}{(M_X-m_{\chi_{c1}})^2+
  \frac{\Gamma^2_{\chi_{c1}}}{4}} \times \frac{1}{1-\Sigma_{c\bar
    c}^\prime(M_X)}\times
\frac{\delta_n^2}{3\pi}E_\gamma^3\frac{M_{\psi(nS)}}{M_X} \label{eq:rad3}
\end{equation}
with $E_\gamma= (M_X^2-M_{\psi(nS)}^2)/(2M_X)$.  The first factor
deviates from one when the width of the dressed $\chi_{c1}(2P)$ starts
growing and becomes comparable with $M_X-\mbareuno$.  The factor  $1/(1-\Sigma_{c\bar
    c}^\prime(M_X))$ is  $\tilde Z_{\X}=1-\tilde X_{\X}$ (see
Eq.~(\ref{eq:defXZ})), and it can be identified with 
the probability to find the compact component
$\chi_{c1}(2P)$ in the physical wave function of the $X(3872)$. On the
other hand, the last factor is
\begin{equation}
\frac{\delta_n^2}{3\pi}E_\gamma^3\frac{M_{\psi(nS)}}{M_X} = \left
\{\begin{array}{cc} 89 ~ {\rm keV}, & 2S \cr
                    60 ~{\rm keV}, & 1S\end{array} \right. \label{eq:rad4}
\end{equation}
using the matrix elements $\delta_{1S}=0.046$ GeV$^{-1}$ and
$\delta_{2S}= 0.38$ GeV$^{-1}$. We have estimated $\delta_{nS}$  from the
widths given in Table III of Ref.~\cite{Barnes:2005pb} for the 2P E1
radiative transitions calculated  with the non-relativistic potential
model. (We have used $M_{J/\psi}= 3096.92$ MeV, $M_{\psi(2S)}=
3686.11$ MeV and the mass predicted in Ref.~\cite{Barnes:2005pb} for
the $\chi_{c1}(2P)$ state.)

The estimate in Eq.~(\ref{eq:rad3}) depends on the renormalization scheme and
should cancel the dependence on scale of the meson loop
contributions. Here, we have computed it using an UV cutoff,
$\Lambda=1$ GeV, while the meson loops were evaluated in ~\cite{Guo:2014taa}
using dimension regularization with the $\overline{{\rm MS}}$ subtraction
scheme at $\mu=M_X/2,M_X,2M_X$. 

We pay attention to the two meson
loop function,  and compare 
$G_{\rm QM}(E)/\left(4M_DM_{D^*}
e^{-k^2/\Lambda^2}\right)$ (Eq.~(\ref{eq:gmat_gr})), with
$G^{\overline{MS}}(s,\mu)$, defined as 
\begin{eqnarray}
G^{\overline{MS}}(s,\mu) &=&
i\int\frac{d^4q}{(2\pi)^4}\frac{1}{q^2-M^2_D}\frac{1}{(P-q)^2-M^2_{D^*}}\nonumber\\
&=&\overline{G}(s) + \frac{1}{16\pi^2}\left\{
-2 +\frac{1}{M_D+M_{D^*}}\left(M_D\log\frac{M_D^2}{\mu^2}+ M_{D^*}\log\frac{M_{D^*}}{\mu^2}\right)\right\}
\end{eqnarray}
with $P^\mu$ the total four momentum ($P^2 = s$), and the finite and
scale independent function $\overline{G}(s)=
G^{\overline{MS}}(s,\mu)-G^{\overline{MS}}(s=(M_D+M_{D^*})^2,\mu)$,
given in Eq. (A9) of Ref.~\cite{Nieves:2001wt}. From such comparison,
and looking at the FRS and in the vicinity of $s=M_X^2$,  we 
find that scales $\mu$ of the other of $M_X$ would correspond to
UV cutoffs, $\Lambda$, much larger than 1 GeV, or equivalently
$\Lambda=1$ GeV would correspond to a $\overline{MS}$ scale $\mu$ of
the order of 1 GeV, significantly smaller than $M_X$. 

\begin{figure}
\begin{center}
\makebox[0pt]{\includegraphics[width=0.4\textwidth]{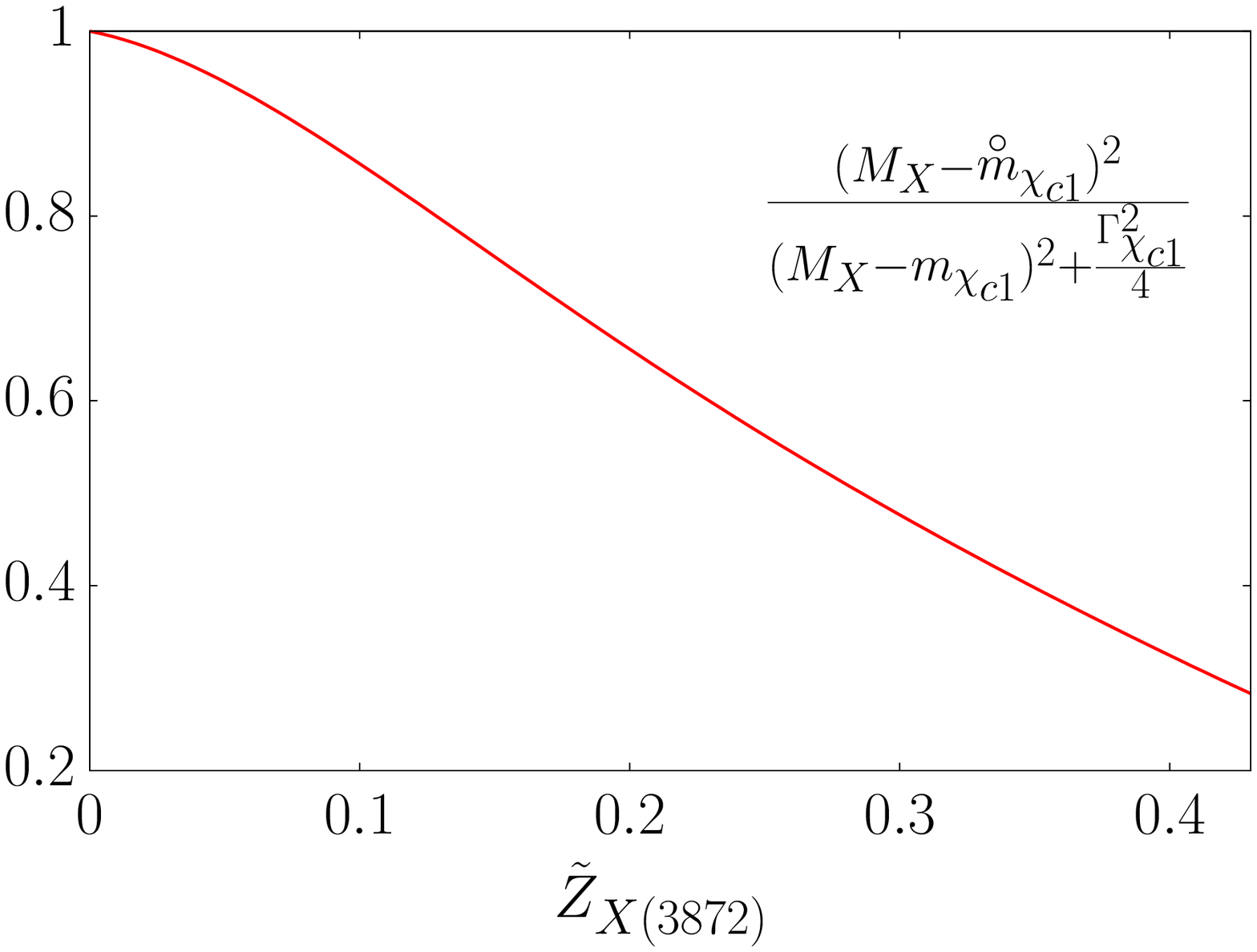}\vspace{0.3cm}\includegraphics[width=0.5\textwidth]{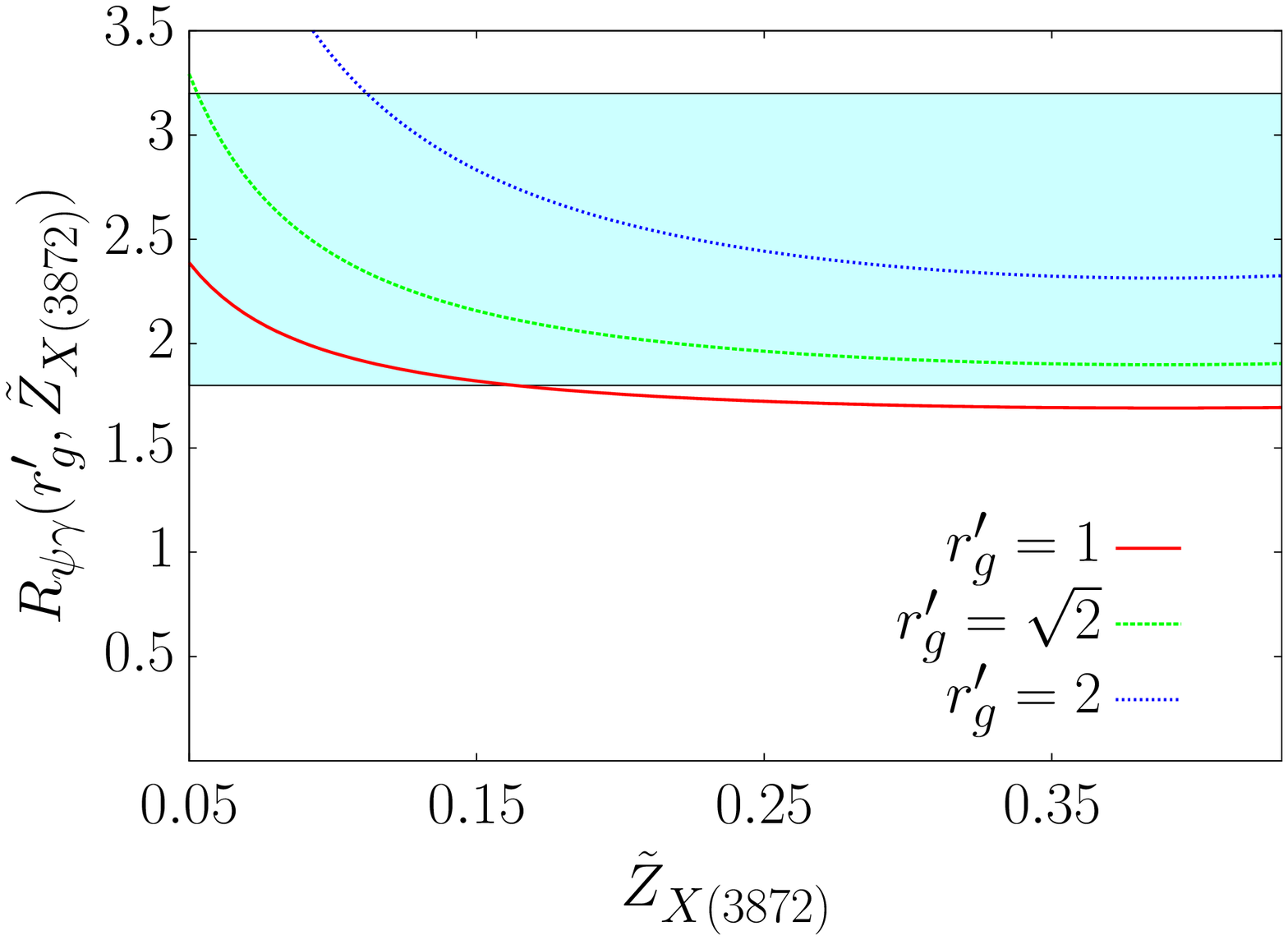}}
\caption{Function $ f(\tilde Z_{\X})$ (left) entering in the
  definition of the ratio $R_{\psi\gamma}(r'_g,\tilde Z_{\X})$  in
  Eq.~(\ref{eq:rad5}). This latter ratio is shown in the right panel for three different
  values of $g'=1,\sqrt{2}$ and $2$ (units of 2 GeV$^{-3/2}$), together with the experimental
  band $R_{\psi\gamma}=2.5 \pm 0.7$ from Ref.~\cite{Aaij:2014ala}. All calculations have been carried out
  with an UV cutoff $\Lambda= 1$ GeV. }\label{fig:rad2}
\end{center}
\end{figure}

We cannot increase the size of the UV cutoff within the EFT proposed 
in \cite{Nieves:2012tt,HidalgoDuque:2012pq} to describe the $X(3872)$,
since we will be  breaking HQSS and our estimate of the counter-term
will not be realistic. However, we can run down the  charmed meson loop contribution to
the radiative decays calculated in  ~\cite{Guo:2014taa} to scales
$\mu\sim $ 1 GeV. In the case of the $\psi(2S)\gamma$ mode such
running seems stable and leads to (Eq.~(\ref{eq:XJpsi2})) 
\begin{equation}
\Gamma^{\rm loops}(\X\to \psi(2S) \gamma) \sim  2.7 (r_xr'_g)^2\,[{\rm
    keV}] ~{\rm at} ~\mu=1~{\rm GeV},
\end{equation}
while we will assume that the hadron loop contribution to the $\X\to J/\psi
\gamma$ decay is much smaller than 1 keV at scales of the order of 1
GeV, as the running in Eq.~(\ref{eq:XJpsi1}) seems to suggest. Thus, we
consider following the discussion in Ref.~\cite{Guo:2014taa} (taking into account also the results of Eqs.~(\ref{eq:rad3})
and (\ref{eq:rad4})),
\begin{eqnarray}
R_{\psi\gamma}(r'_g,\tilde Z_{\X}) &=&\left.\frac{B_{r}(X\to\psi(2S)\gamma)}{B_{r}(X\to J/ \psi
  \gamma)}\right|_{\rm
  loops~+~counter-term~of~Fig.~\ref{fig:rad}}\nonumber\\
& \sim &
\frac{70\tilde Z_{\X}\times f(\tilde Z_{\X}) +(1-\tilde Z_{\X}) 2.7
  r^{\prime\, 2}_g}{56\tilde Z_{\X}\times f(\tilde Z_{\X})} \label{eq:rad5}
\end{eqnarray}
where $ f(\tilde Z_{\X})$ (shown in the left panel of
Fig.~\ref{fig:rad2}) accounts for the dressed and bare charmonium
propagator ratio squared that appear in Eq.~(\ref{eq:rad3}). The above
approximation for $R_{\psi\gamma}$ only makes sense as long as $\tilde
Z_{\X}$ is larger than let us say 0.05 to justify having neglected the
meson loop contribution in the $\X\to J/\psi \gamma$ mode. We have
also neglected any correction due to an imprecise knowledge of the
$XD\bar D^*$ coupling, $r_x$, and more importantly to possible
destructive or constructive interferences between the meson-loops and
the counter-term (quark-loops) contributions in the $\psi(2S)\gamma$
decay. We are aware these latter effects might be
important~\cite{Dong:2009uf}, but we cannot properly estimate them in
this exploratory study, where we aim at discussing the implications of
the existence of quarkonium components in the $X(3872)$ in the
dynamics of the predicted $X_2(4012)$ resonance, as well as in the
properties of the possible partners of these charmed resonances in the
bottom sector. Note that the sign of $g'$ is uncertain, which is also
a limitation for the scheme of Ref.~\cite{Dong:2009uf}. Moreover, we
should also acknowledge that the counter-term needed in
\cite{Guo:2014taa} might involve contributions for other type of
short-range physics, as for instance higher momentum components of the
hadronic $X(3872)$ wave function. Thus, the discussion below can only be
qualitative. 

In the right panel of
Fig.~\ref{fig:rad2}, the ratio $R_{\psi\gamma}(r'_g,\tilde Z_{\X})$ is
shown as a function of $Z_{\X}$ for three different values of the
$\psi(2S) D^{(*)} \bar D^{(*)}$ coupling constant, together with the
experimental band given in Eq.~(\ref{eq:Rexprad}) (we have added in
quadratures statistical and systematic errors).
From Fig.~\ref{fig:rad2}, we conclude that moderate $\X$ charmonium
contents in the range $\tilde Z_{\X}=0.1-0.3$ lead to successful
descriptions of the $R_{\psi\gamma}$   considering ratios $g'/g >
1$ in line with the expectations of Ref.~\cite{Dong:2009uf}. Indeed,
if this ratio is of the order of 2, larger $\X$ charmonium
contents can be easily accommodated, though in that case the 
experimental ratio of decay fractions of $X(3872)$ into 
$J/\psi\pi^{+}\pi^{-}$ and $J/\psi\pi^{+}\pi^{-}\pi^{0}$ final states might be 
difficult to be explained. 

From the results in Table~\ref{tab:dvsX}, and bearing in mind all sort
of shortcomings mentioned above,  we expect the mixing parameter
$d(\Lambda=1\,{\rm GeV})$ to lie in the 0.1 -- 0.25
fm$^{1/2}$ interval, which would correspond to $X(3872)$ meson-molecular
probabilities in the  0.9 -- 0.65 range.

\subsubsection{Discussion}
From the above considerations, the dressed charmonium state
$\chi_{c1}(2P)$ should have a mass around 3910 -- 3925 MeV, with a
width in the range 5 -- 70 MeV and a sizable molecular ($D\bar D^*$)
component, in the interval 6-40\%, depending on the specific value of
$d$ (see Tables~\ref{tab:dvsX} and \ref{tab:500charm}). These results
are similar to those found in the quark model of
Ref.~\cite{Ortega:2010qq}, where charmonium and $D\bar D^*$
configurations are coupled using the $^3P_0$ approximation. There, the
elusive $X(3872)$ meson appears as a new state with a high probability
for the $D\bar D^*$ molecular configuration, and a sizable $c\bar c\,
2^3P_1$ component (7 -- 30\% depending on the strength of the used
$^3P_0$ interaction). The original $\chi_{c1}(2P)$ state acquires also
a sizable meson molecular content (10 -- 20\%), and it is identified
in \cite{Ortega:2010qq} with the $X(3940)$, whose PDG mass and width
are~\cite{Agashe:2014kda} $3942\pm 9$ MeV and $37^{+27}_{-17}$ MeV,
respectively. Our predicted width for the charmonium dressed state is
in good agreement with that of the $X(3940)$, though the mass is
somehow low. The mass of the bare $c\bar c\, 2^3P_1$ state used in
\cite{Ortega:2010qq} is significantly larger (3947.4 MeV) than that
used here (3906 MeV), which brings the mass of the dress charmonium
state in \cite{Ortega:2010qq} naturally closer to that of the
$X(3940)$ resonance. Note however, that neither the width of the
dressed $c\bar c\, 2^3P_1$ state nor the ratio $R_{\psi\gamma}$ of
$X(3872)$ radiative decays are calculated in
\cite{Ortega:2010qq}. Moreover, within the approach of this latter
reference the meson loops slightly decrease the mass of the charmonium state,
opposite to what we find in this work.

The phenomenological work of Ref.~\cite{Dong:2009uf} relies in the
inspired quark model findings of  Ref.~\cite{Swanson:2003tb} to
quantify the molecular components of the $X(3872)$, while the interplay
between   its charmonium and 
molecular components is determined from the ratio $R_{\psi\gamma}$ of
radiative decays, as we have qualitatively done here. The findings of
Ref.~\cite{Dong:2009uf} favor an admixture of 5 -- 12\% of a $\bar c c$
component, which  can be easily accommodated  within our results.

Thus, our results together with those of
Refs.~\cite{Dong:2009uf,Ortega:2010qq} do not support other
interpretations of the $X(3872)$, for instance that of
Ref.~\cite{Ferretti:2014xqa}, where this resonance is described as a $c\bar c$ core
plus higher Fock components due to the coupling to the 
meson-meson continuum, which is thought to be compatible with the
meson $\chi_{c1}(2P)$. 


\subsection{Numerical results: the $2^{++}$ hidden charm sector}
\label{sec:2++charm}

\begin{table}[h]
\begin{tabular}{cc|ccc|ccc}
$d$ & 
  $\tilde X_{X(3872)}$ & $g^{\chi_{c2}}_{D^*\bar D^*}$ &   $\tilde
  X_{\chi_{c2}}$ & $\mbaredos $  & $M_{X_2}-2 M_{D^*}-i\frac{\Gamma_{X_2}}{2}$ & $g^{X_2}_{D^*\bar
    D^*}$ &  $\tilde X_{X_2}$ \\
$[{\rm fm}^{1/2}]$ &  & $[{\rm GeV}^{-1/2}]$ 
  & &  ${\rm [MeV]}$ & ${\rm [MeV]}$& $[{\rm GeV}^{-1/2}]$ &  \\\hline
0.& 1 &0.0 & 0.0 & 3927.2 & $-5.6$ & 0.97 & 1. \\ 
0.05 & 0.98 & 0.27 &  0.01 & 3927.8 & $-4.5$    & 0.90 & 0.996 \\
0.10 & 0.92&  0.51 &  0.02 & 3929.6 & $-1.8$    & 0.67 & 0.991 \\
0.15 & 0.84& 0.69 &  0.04 & 3932.2 & $-0.0$ at SRS & $-0.12\, i$ & $> 1$
\\
0.20 & 0.75& 0.82 &  0.05 & 3935.2 & $-6.4$ at SRS & $-0.76\, i$ & $> 1$  \\
0.22 & 0.71& 0.86 &  0.06 & 3936.4 & $-21.2$ at SRS & $-1.24\, i$ & $> 1$
\\
0.25 & 0.66& 0.90 &  0.06 & 3938.3 & $-28.3- \frac{72.9}{2}\,i$   &
$0.23-0.65\,i$~~ & $0.47+0.32\,i$  \\
0.30 & 0.57& 0.95 &  0.07 & 3941.2 & $-31.2- \frac{162.8}{2}\,i$   &
$0.03+0.67\,i$~~ & $0.48-0.04\,i$  \\
0.35 & 0.49& 0.96 &  0.07 & 3943.8 & $-59.5- \frac{312.6}{2}\,i$   & $0.30+0.71\,i$~~ & $0.52-0.39\,i$ \\\hline
\end{tabular}
\caption{ Properties of the $2^{++}$
hidden charm poles as a function of $d$.  We solve  Eq.~(\ref{eq:pole-position})
with $\Lambda= 1.0$ GeV and $C_{0X}(d)$, determined from
Eq.~(\ref{eq:defc0x}), can be found in Table~\ref{tab:dvsX}. The position of the dressed $\chi_{c2}(2P)$ is
fixed at $m_{\chi_{c2}}^{\rm exp}=3927.2$ MeV in the FRS, and we also
give the $X(3872)$ meson-molecular
probabilities  ($\tilde X_{X(3872)}$) for each value of $d$. }\label{tab:dvsX2}
\end{table}
\begin{figure}[b]
\begin{center}
\makebox[0pt]{\includegraphics[width=0.32\textwidth]{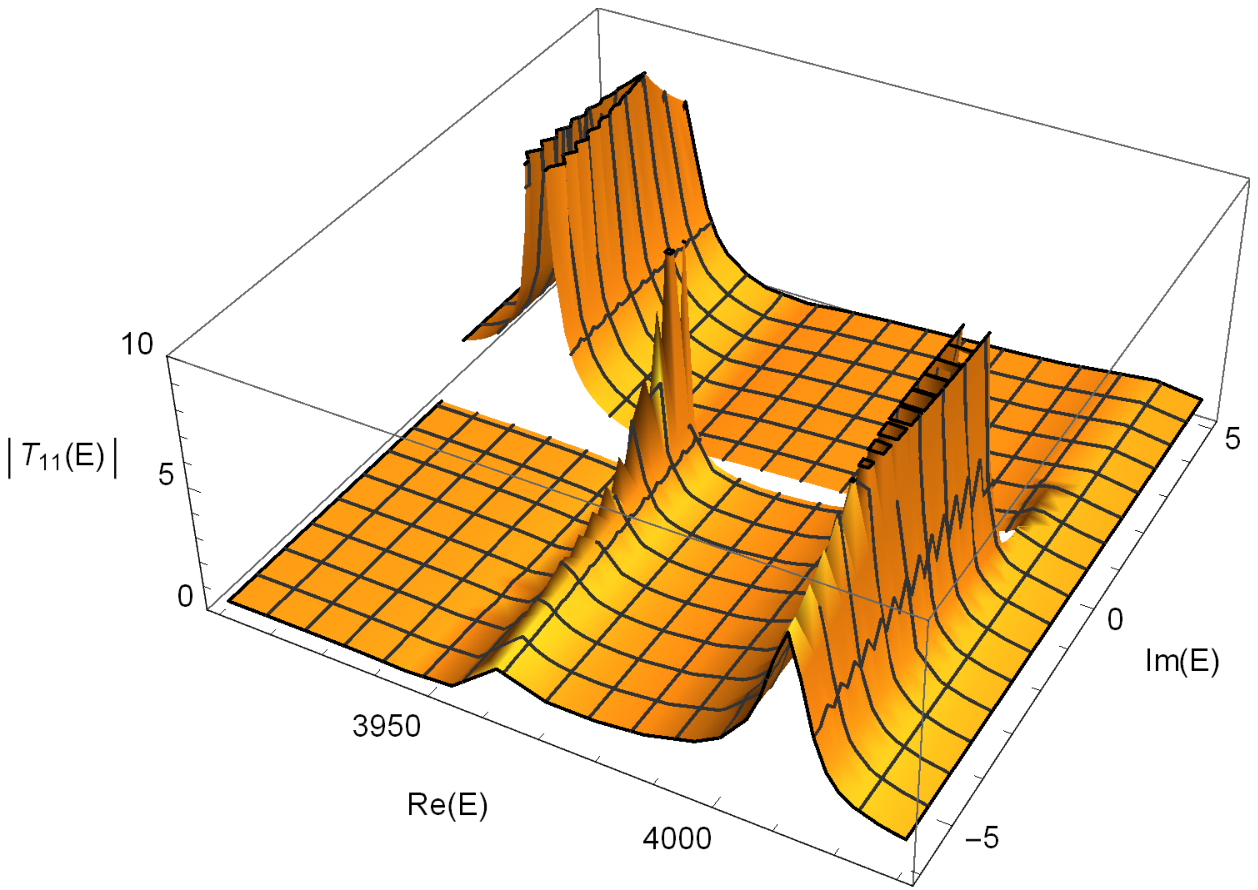}\hspace{0.35cm}\includegraphics[width=0.32\textwidth]{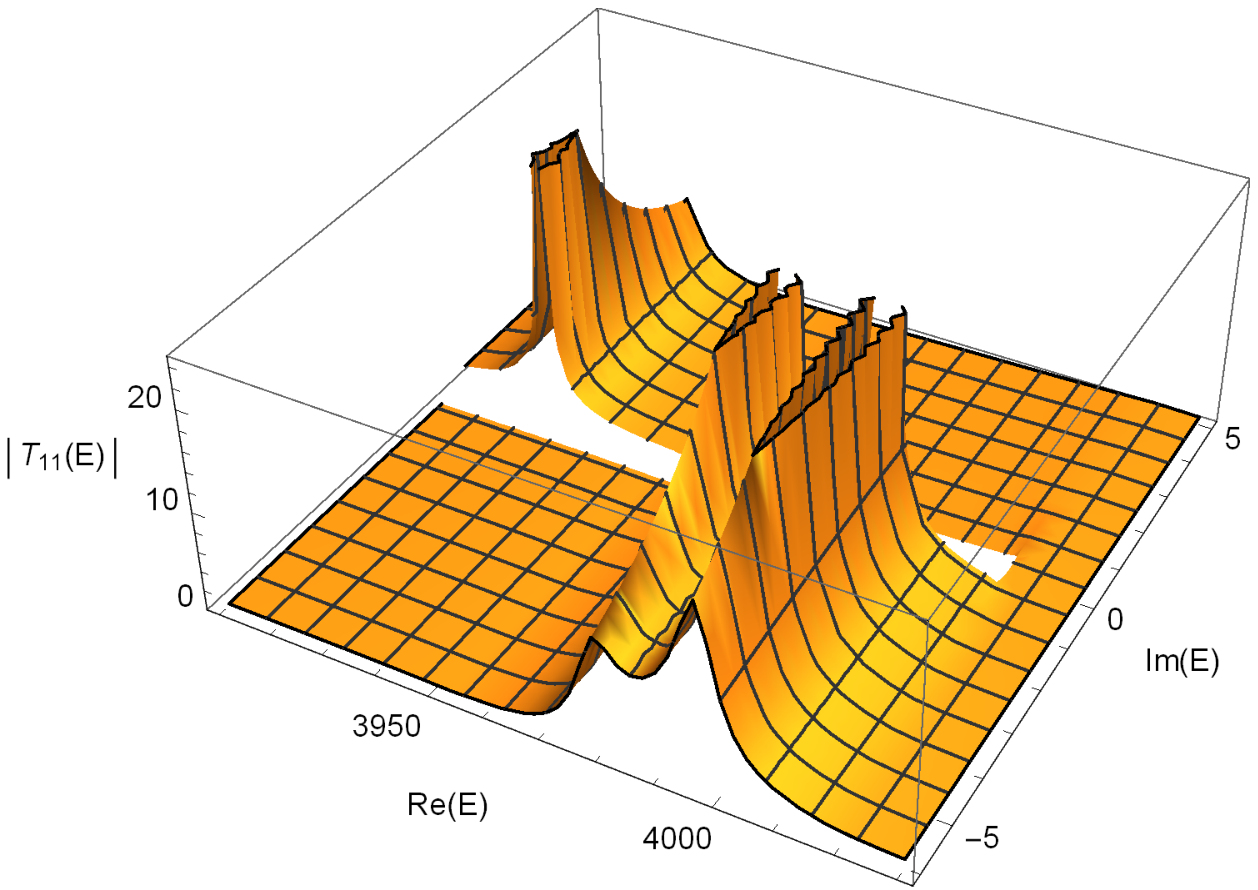}\hspace{0.35cm}\includegraphics[width=0.32\textwidth]{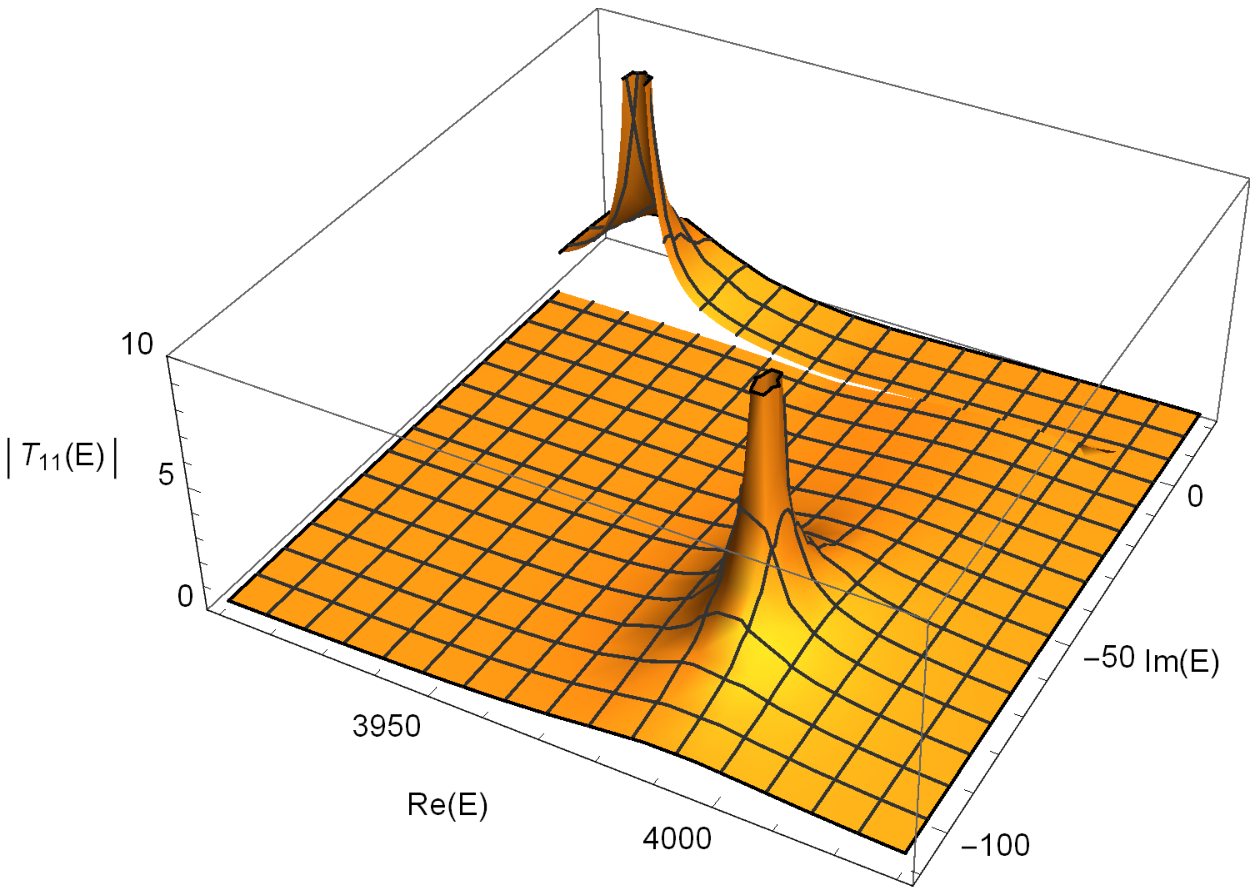}} 
\caption{Hidden charm $J^{PC}=2^{++}$ sector. FRS (${\rm Im}(E)> 0$)
  and SRS (${\rm Im}(E)< 0$) of $|T_{11}(E)|$ [fm$^2$]
  (Eq.~(\ref{eq:defT11})) as a function of the complex energy $E$
  [MeV], for $d=0.20$ (left), 0.22 (middle) and 0.25 (right)
  fm$^{1/2}$. Note that, since the $T-$matrix is shown for only half
  of the SRS (and also the FRS), the pole in the SRS conjugate to the
  pole shown in the figures is not visible. In the first two plots,
  there appear one pole in the FRS ($\chi_{c2}(2P)$) located at 3927.2
  MeV and two more in the real axis of the SRS below threshold and
  disconnected from the FRS. In the left (middle) plot, the pole
  located at 4010.9 (3996.0) MeV would correspond to the $X_2(4012)$
  (HQSS partner of the $X(3872)$) state, while the other one, located
  at 3959.5 (3978.1) MeV, arises because of the bare $\chi_{c2}$ pole
  included in the amplitudes. Finally in the right plot, there are
  appear the FRS $\chi_{c2}(2P)$ pole and a second one deep into the
  SRS complex plane. All calculations have been carried out with an UV
  cutoff $\Lambda= 1$ GeV.  The ``serrated'' appearance of the poles
  in the first plot is due to the coarse mesh used to create the
  surface plot. It can be eliminated by using a finer mesh, which
  would require the computation of the amplitude for a larger number
  of complex energies. }\label{fig:2++}
\end{center}
\end{figure}
The effective interactions in the $1^{++}$ and $2^{++}$ sectors
at the $X(3872)$ mass and the $D^*\bar D^*$ threshold, 
are:
\begin{eqnarray}
V_{\rm eff}^{1^{++}}(E=M_X) &=& C_{0X}
+\frac{d^{\,2}}{M_X-\mbareuno}=\frac{1}{G_{\rm QM}(M_X)} \\ 
V_{\rm  eff}^{2^{++}}(E=2M_{D^*}) &=&  C_{0X}
+\frac{d^{\,2}}{2M_{D^*}-\mbaredos} \nonumber\\
 &=& V_{\rm eff}^{1^{++}}(E=M_X) +d^{\,2}\left(
\frac{(2M_{D^*}-M_X)-(\mbaredos-\mbareuno)}{(2M_{D^*}-\mbaredos)(\mbareuno-M_X)}\right)
\end{eqnarray}
and hence $V_{\rm eff}^{2^{++}}(E)-V_{\rm
  eff}^{1^{++}}(E=M_X) > 0$, for $E$ in the vicinity of the
$D^*\bar D^*$ threshold, because we expect $(2M_{D^*}-M_X)\sim m_\pi
>(\mbaredos-\mbareuno)$. Indeed, for $d=d^{\rm
  \,crit}$, $C_{0X}=0$, and thus the net interaction in the $2^{++}$ sector
will be repulsive since $2M_{D^*} > \mbaredos$.

In what follows, we will fix $\mbaredos$ such that the dressed $2P$
quarkonium mass ($ m_{\chi_{c2}}$) will be equal to
$m_{\chi_{c2}}^{\rm exp}$. In Table~\ref{tab:dvsX2}, we show the
properties of the poles found in the $2^{++}$ hidden charm sector as a
function of the mixing LEC $d$. We solve Eq.~(\ref{eq:pole-position})
with an UV cutoff of 1 GeV as in the case of
Table~\ref{tab:dvsX}. First, we see that $\mbaredos$ and
$m_{\chi_{c2}}^{\rm exp} $ differ just in few MeVs, and hence we check
the $D^*\bar D^*$ loops have little influence on the charmonium level,
though it develops a sizable coupling to the meson pair. Moreover
$\mbaredos > m_{\chi_{c2}}^{\rm exp}$, since $\Sigma_{c\bar
  c}(m_{\chi_{c2}}^{\rm exp}) <0$ in the FRS and for this regime of
$C_{0X}$ values and energies. As $d$ increases, the molecular
$X_2(4012)$ (HQSS partner of the $X(3872)$) state approaches to
$2M_{D^*}$, and for $d> 0.15$ fm$^{1/2}$ it crosses to the SRS, moving
quickly away from threshold along the real axis\footnote{Note that
  $\Sigma_{c\bar c}(E) >0$ in the SRS, for real energies below
  $2M_{D^*}$ and $d$ around 0.15 fm$^{1/2}$ because the loop factor
  $(1-C_{0X}G_{\rm QM}^{II})$ takes negative values}.  Actually, what
happens is that the $X_2(4012)$ pole at the SRS merges with a replica
of the bare $\chi_{c2}(2P)$ pole, as illustrated in
Fig.~\ref{fig:2++}, and the new pole gets deep into the complex plane
when $d$ increases above 0.22 fm$^{1/2}$.

From the above discussion on the $X(3872)$ radiative decays, we expect
the mixing LEC $d$ to take values in the range 0.1 -- 0.25 fm$^{1/2}$
for $\Lambda=1$ GeV, which in turn would imply that the $X_2(4012)$
would likely lie in the SRS, below threshold disconnected from the
FRS, either in the real axis or deep into the complex plane. Note
 that for values of $d$ close to $d\simeq 0.15$ fm$^{1/2}$,
even in cases where the pole is in the SRS below threshold, it could
however have sizable effects on the observables, since it would be
close to the $D^*\bar D^*$ threshold, where SRS and FRS are
connected. Considering equivalent molecular components of the
$X(3872)$, the conclusions obtained with $\Lambda=0.5$ GeV are
qualitatively similar, as can be seen in
Table~\ref{tab:500charm}\footnote{The $\Lambda=0.5$ and $\Lambda=1$
  GeV $X_2$ predicted masses, calculated neglecting the quarkonium
  mixing ($d=0$) , are similar (they differ by less than 1 MeV) and
  for $d=0$ the $X_2$ state would be located around 5 MeV below the
  $D^*\bar D^*$ threshold.  The $\chi_{c2}(2P)$ is much lighter,
  around 85 -- 90 MeV, and in this case the form-factor $f_\Lambda$
  that appears in Eq.~(\ref{eq:defg1}) is around twice larger for
  $\Lambda=0.5$ GeV than for $\Lambda=1$ GeV. We see this dependence
  on the UV cutoff in $g_{D^*\bar D^*}^{\chi_{c2}}$, coupling of the
  $\chi_{c2}(2P)$ state to the $D^*\bar D^*$ meson pair, which for
  similar molecular components of the $X(3872)$ is around 2 -- 3 times
  larger for $\Lambda=0.5$ GeV than when it is calculated using
  $\Lambda=1$ GeV, reflecting a large off-shell ambiguity for this
  coupling. This cutoff dependence cancels out for instance in the
  completeness relation of Eq.~(\ref{eq:compo}) or in the relation
  among quarkonium and meson--molecular couplings of
  Eq.~(\ref{eq:g2vsg1}).}.

Thus, the different interplay of the charmonium components in the $X(3872)$ and in its
hypothetical $2^{++}$ HQSS partner makes plausible that
this latter state is not accessible to the direct observation, or in
other words, that it does not exist as an actual QCD
state\footnote{This is somehow an abuse of
language. We call "actual QCD states" as states that produce observable
effects. If a SRS pole is located below threshold but 
deep in the complex plane, or it is close to the
real axis, but much below the
threshold, it will not produce any observable effects, and hence it
will  be impossible to detect. }. Within the model
developed in Ref.~\cite{Ortega:2010qq}, it is also found insufficient
attraction in the $2^{++}$ sector to create an additional, mostly
$D^*\bar D^*$ molecular, state~\cite{Entem:2016ojz}. Moreover, we
should remind here that in the scheme of Ref.~\cite{Baru:2016iwj},
mass and width of this state were strongly affected by the one-pion
exchange interaction in coupled channels. 

\newpage

This state in the $2^{++}$ sector was predicted in
\cite{Nieves:2012tt,HidalgoDuque:2012pq,Guo:2013sya}, where it was
also shown that even considering 15-20\% HQSS violations its existence
seemed to be granted. However, the $X_2(4012)$ has not been observed
yet, and hence the study carried out here might shed light into this
issue. This also shows that corrections stemming from
charmonium admixture in the molecular $X(3872)$, enhanced/distorted by
threshold effects, need to be explicitly considered exhibiting their
energy dependence, and they cannot be
just accounted for in the short-distance meson-meson LECs.

\subsection{Numerical results: the hidden bottom $1^{++}$ and $2^{++}$ sectors.}
\label{sec:bottom}
\begin{table}
\begin{center}
\begin{tabular}{ccc|ccc}
\multicolumn{3}{c|}{$1^{++}$}&\multicolumn{3}{c}{$2^{++}$}\\
state & mass  & $B\bar B^*$ threshold & state & mass & $B^*\bar B^*$ threshold\\\hline
$X_b$ ($\Lambda=1$ GeV)~\cite{Guo:2013sya} & $10539^{+25}_{-27}$ & 10604.2 & $X_{b2}$($\Lambda=1$ GeV)~\cite{Guo:2013sya}& $10584^{+25}_{-27}$
& 10649.7 \\
$X_b$ ($\Lambda=0.5$ GeV)~\cite{Guo:2013sya} & $10580^{+9}_{-8}$ &  & $X_{b2}$($\Lambda=0.5$ GeV)~\cite{Guo:2013sya}& $10626^{+8}_{-9}$
& \\\hline
$\chi_{b1}(1P)$ & $9892.78 \pm 0.40$ & & $\chi_{b2}(1P)$ & $9912.21 \pm
0.40$ \\
$\chi_{b1}(2P)$ & $10255.46 \pm 0.55$ & & $\chi_{b2}(2P)$ & $10268.65 \pm
0.55$ \\
$\chi_{b1}(3P)$ & $10512.1 \pm 2.3$ & & $\chi_{b2}(3P)$ & $10522.1^\dagger$ \\\hline
\end{tabular}
\end{center}
\caption{Masses of several hidden bottom states and thresholds in
  MeV. We use the isospin averaged $B-$meson mass, $M_B=5279.40$ MeV,
  and for the vector meson we take $M_{B^*}= 5324.83$
  MeV~\cite{Agashe:2014kda}. The $X_b$ and $X_{b2}$ are heavy-quark
  spin-flavor partners of the $X(3872)$ predicted in
  \cite{Guo:2013sya}. We quote here the masses found in this reference
  for $\Lambda=$1 and 0.5 GeV, and the errors account for heavy quark symmetry
  breaking
  corrections. The masses of the $\chi_{bJ}(nP)$ are taken from the
  PDG~\cite{Agashe:2014kda}, with errors added in
  quadratures, except for that of the $2^{++}$ 3P state. $\dagger:$ Theory
predictions for the $\chi_{b2}(3P)-\chi_{b1}(3P)$ mass splitting vary from 8
to 12 MeV~\cite{Kwong:1988ae,Motyka:1997di,Segovia:2016xqb}. We set here this splitting  to 10 MeV.}\label{tab:masses-bottom}
\end{table}
\begin{table}[b]
\begin{tabular}{cc|ccc|ccc}
$d$  & 
  $\tilde X_{X(3872)}$ & $g^{\chi_{b1}}_{B\bar B^*}$ &   $\tilde
  X_{\chi_{b1}}$ & $\mbareunob $  & $E_{X_b}-M_B-M_{B^*}$ & $g^{X_b}_{B\bar
    B^*}$ &  $\tilde X_{X_b}$ \\
$[{\rm fm}^{1/2}]$ & & $[{\rm GeV}^{-1/2}]$ 
  & &  ${\rm [MeV]}$ & ${\rm [MeV]}$& $[{\rm GeV}^{-1/2}]$ &  \\\hline
0. & 1& 0.0 & 0.0 & 10512.1 & $-65.9$ &2.30  & 1. \\ 
0.05 &0.98 &0.98 & 0.09 & 10515.0 & $-60.7$ &2.04  & 0.91 \\ 
0.10 &0.92 &1.46 & 0.20 & 10521.4 & $-47.6$ &1.55  & 0.80 \\ 
0.15 &0.84 &1.59 & 0.24 & 10527.8 & $-30.8$ &1.11  & 0.77 \\ 
0.20 & 0.75&1.57 & 0.23 & 10532.6 & $-13.1$ &0.69  & 0.80 \\ 
0.25 & 0.66&1.49 & 0.21 & 10536.1 & $-0.1$ &0.16  & 0.96 \\ 
0.30 & 0.57&1.40 & 0.18 & 10538.5 & $4.9- \frac{68.2}{2}\,i$ &
$0.05-0.26\,i$~~ &  $0.43+0.16\,i$ \\ 
0.35 & 0.49&1.29 & 0.16 & 10540.2 & $44.8- \frac{181.4}{2}\,i$ & $0.12+0.28\,i$~~ &  $0.55-0.21\,i$ \\\hline
\end{tabular}
\caption{ Properties of the $1^{++}$
hidden bottom poles as a function of $d$.  We solve  Eq.~(\ref{eq:pole-position})
with $\Lambda= 1.0$ GeV and $C_{0X}(d)$, determined from
Eq.~(\ref{eq:defc0x}), can be found in Table~\ref{tab:dvsX}. The position of the dressed $\chi_{b1}(3P)$ is
fixed at $m_{\chi_{b1}}^{\rm exp}=10512.1$ MeV in the FRS, and we also
give the $X(3872)$ meson-molecular
probabilities  ($\tilde X_{X(3872)}$) for each value of $d$.  }\label{tab:dvsXb}
\end{table}
In Table~\ref{tab:masses-bottom}, we compile the masses of the
bottomonium states quoted in the PDG in the $1^{++}$ and $2^{++}$
sectors, together with those of the hidden bottom 
partners of the $X(3872)$ and the $X_2(4012)$ predicted in \cite{Guo:2013sya}. As we warned the
reader in the introduction, the bottom and charm sectors were
connected in \cite{Guo:2013sya} by assuming the bare couplings in the
$4H$ interaction Lagrangian of Eq.~(\ref{eq:LaLO}) to be independent
of the heavy quark mass. Neither the $X_b$, nor the $X_{b2}$ have been observed
yet, as it happens for the $X_2(4012)$. Moreover their predicted masses
show an important UV cutoff dependence.   We first focus on the
$\Lambda=1$ GeV case  because for this value of the UV cutoff, the
predicted binding energies of both $X_b$ and $X_{b2}$  are much larger than those obtained in the
$\Lambda=0.5$ GeV case ($\simeq$ 65 MeV versus $\simeq$ 25
MeV). Nevertheless results for this latter UV cutoff can be found in
the Appendix, and will be considered  for the general discussion.

We fix $\mbareunob$ and $\mbaredosb$ by requiring that the dressed
quarkonium masses $ m_{\chi_{bJ}}$ will match those of the $3P$ states
quoted in Table~\ref{tab:masses-bottom}. The bare states lie below the
$X_b$ and $X_{b2}$ states, which produces some repulsion, as in the
case of the hidden charm $X_2$ state. Constituent quark models predict
additional bottomonium states. Here, we pay attention to the spectrum
obtained in the recent work of Ref.~\cite{Segovia:2016xqb}, where the
non-relativistic $Q\bar Q$ interaction used in
Ref.~\cite{Ortega:2010qq} is employed and a global agreement with the
experimental pattern is found.  Among the higher levels reported in
~\cite{Segovia:2016xqb}, the $4\,^3P_1(10737)$, $2\,^3F_2(10569)$,
$4\,^3P_2(10744)$ and $3\,^3F_2(10782)$ might have some relevance for
the present discussion~\cite{Entem:2016ojz}. The $4P$ states are
heavier than the $X_b$ and $X_{b2}$, and are located around 130 and 95
MeV above the $B\bar B^*$ and $B^*\bar B^*$ thresholds,
respectively. These levels would produce extra attractions. On the
other hand and because of the large orbital angular momentum, the
$F-$states in the $2^{++}$ sector seem to play a really sub-dominant
role~\cite{Entem:2016ojz} in the dynamics of the $X_{b2}$. We will
examine here the worst of the scenario for the existence of the $X_b$
and $X_{b2}$ states, and we will consider only the $3P$ states,
neglecting any attraction from the $4P$ bottomonia.

The contact interaction term $C_{0X}$ is fixed from the $X(3872)$ mass, and
thus its magnitude depends on the LEC $d$ that mixes the molecular
$D\bar D^*$ and $\chi_{c1}(2P)$ components. The presence of the
charmonium state provides an effective attraction that 
contributes to bind the $X(3872)$, which translates in a smaller
$|C_{0X}|$, as seen in Tables~\ref{tab:dvsX} and
\ref{tab:500charm} for $\Lambda=1$ and $0.5$ GeV,
respectively. Assuming the same value for $C_{0X}$ in the bottom
sector, we still need to determine the mixing parameter in the bottom
sector ($d^{\rm bottom}$), which  as discussed in
Subsect.~\ref{sec:bottom} depends in principle on the heavy quark flavor.
Through this LEC, the $3P$ bottomonium states will produce some repulsion in
the effective $B^{(*)}\bar B^{(*)}$ interaction. 

For practical purposes, we will assume the mixing of molecular and quarkonium
components independent of both flavor and the $Q\bar Q$
radial\footnote{Note that in charmonium, we considered the $c\bar c$
  pair in the $2P$ wave, while in bottomonium, the $3P-$states would
  be the closest ones to the $X_b$ and $X_{b2}$ resonances.  } quantum
number in the heavy quark limit. Even if inexact, these assumptions
will allow us, at least qualitatively, to obtain an idea on the
effects of quarkonium--molecular configurations admixtures on the
$X_b$ and $X_{b2}$ states. Thus and from the discussion in
Subsect.~\ref{sec:rad-decays}, we consider the $C_{0X}$ values fixed
from the $X(3872)$, and associated to $d(\Lambda=1\,{\rm GeV})$ in the
range 0.1 -- 0.25 fm$^{1/2}$, and use the same values for $d^{\rm
  bottom}(\Lambda=1\,{\rm GeV})$ to take into account the
repulsion induced by the $\chi_{b1}(3P)$ and $\chi_{b2}(3P)$ states.
Pole positions calculated using $\Lambda = 1$ GeV and different values
of the mixing parameter $d$ are presented in Tables~\ref{tab:dvsXb}
and ~\ref{tab:dvsXb2} for the $1^{++}$ and $2^{++}$ sectors,
respectively. The $d-$dependence is quite similar in both sectors and
it is mostly dictated by the proximity of the resonances to the bottomonium
levels.  We find moderate bare--dressed quarkonium mass differences of
the order 5 -- 25 [5 -- 20] MeV, and molecular meson contents in the
dressed state ranging in the interval 10 -- 20\% [5 -- 10\%] for the
$\chi_{b1}(3P)$ $\left[\chi_{b2}(3P)\right]$ state. On the other hand,
we see that as long the $X(3872)$ meson-molecular component is larger
than 65\% ($\tilde Z_{\X} < 35\%$), both the $X_b$ and $X_{b2}$ states
should exist and should be observed in future experiments.  However,
the different interplay of the quarkonium components in the $X(3872)$
and in its hypothetical $1^{++}$ and $2^{++}$ hidden bottom partners
produces significant changes in the masses of the latter states. Thus,
instead of bindings of the order of 65 MeV, we would expect the
molecular bottom states to lie still below, but much closer to their
respective two meson thresholds, about 45 -- 50 MeV at
most\footnote{The heavy quark symmetry breaking uncertainties quoted
  in Table~\ref{tab:masses-bottom} for these states would account in
  great extent for the changes induced by charmonium contents of the
  $X(3872)$ smaller than 10 -- 15\%.}. Indeed, for the largest
considered admixtures, $d(\Lambda=1\,{\rm GeV})$= 0.2 -- 0.25
fm$^{1/2}$, the $X_b$ and $X_{b2}$ could have binding energies of only
few MeV or less.
\begin{table}[b]
\begin{tabular}{cc|ccc|ccc}
$d$&$\tilde X_{X(3872)}$  & $g^{\chi_{b2}}_{B^*\bar B^*}$ &   $\tilde
  X_{\chi_{b2}}$ & $\mbaredosb $  & $E_{X_{b2}}-2M_{B^*}$ & $g^{X_{b2}}_{B^*\bar
    B^*}$ &  $\tilde X_{X_{b2}}$ \\
$[{\rm fm}^{1/2}]$ & &  $[{\rm GeV}^{-1/2}]$ 
  & &  ${\rm [MeV]}$ & ${\rm [MeV]}$& $[{\rm GeV}^{-1/2}]$ &  \\\hline
0. & 1& 0.0 & 0.0 & 10522.1 & $-66.2$ &2.31  & 1. \\ 
0.05 &0.98 & 0.69 & 0.02 & 10523.4 & $-62.5$ &2.17  & 0.98 \\ 
0.10 & 0.92& 1.20 & 0.06 & 10526.9 & $-52.3$ &1.82  & 0.94 \\ 
0.15 & 0.84& 1.50 & 0.10 & 10531.3 & $-37.2$ &1.37  & 0.91 \\ 
0.20 & 0.75& 1.64 & 0.11 & 10535.7 & $-19.4$ &0.90  & 0.90 \\ 
0.25 & 0.66& 1.67 & 0.12 & 10539.5 & $-3.1$ &0.41  & 0.93 \\ 
0.30 & 0.57&1.64 & 0.11 & 10542.5 & $-18.0- \frac{37.4}{2}\,i$
&$0.16-0.28\,i$~~  & $0.46+0.75\,i$ \\ 
0.35 & 0.49&1.59 & 0.11 & 10545.0 & $\phantom{-}27.1- \frac{195.1}{2}\,i$~~ &$0.09+0.28\,i$~~  & $0.57-0.15\,i$ \\\hline
\end{tabular}
\caption{ Properties of the $2^{++}$
hidden bottom poles as a function of $d$.  We solve  Eq.~(\ref{eq:pole-position})
with $\Lambda= 1.0$ GeV and $C_{0X}(d)$, determined from
Eq.~(\ref{eq:defc0x}), can be found in Table~\ref{tab:dvsX}. 
The position of the dressed $\chi_{b2}(3P)$ is
fixed at $m_{\chi_{b2}}^{\rm exp}=10522.1$ MeV in the FRS, and we also
give the $X(3872)$ meson-molecular
probabilities  ($\tilde X_{X(3872)}$) for each value of $d$.  }\label{tab:dvsXb2}
\end{table}

\newpage

Results obtained using $\Lambda = 0.5$ GeV are presented in the
Table~\ref{tab:500bottom} of the Appendix. Besides the trivial
dependence of the mixing parameter $d$, and of $g^{\chi_{b1}}_{B\bar
  B^*}$ and $g^{\chi_{b2}}_{B\bar B^*}$ on the UV cutoff\footnote{In
  the case of the couplings, it is mostly due to the factor
  $f_\Lambda$ that appears in their definition in
  Eq.~(\ref{eq:defg1}), as we already discussed for the case of the
  $\chi_{c2}(2P)$. 
Indeed in the hidden bottom sector, the
  quarkonium $b\bar b\, 3^3P_{1,2}$ states are located well below
  ($\simeq 90$ and $130$ MeV, respectively) their respective two meson
  thresholds, and $f_\Lambda$ induces a large dependence of the
  couplings on $\Lambda$, around a factor of 4 in the $1^{++}$ sector
  and of 8 in the $2^{++}$ one. }, the conclusions are qualitatively
similar to those discussed above in the $\Lambda = 1$ GeV case. Thus,  
we find also now moderate bare--dressed quarkonium mass differences,
though smaller than for $\Lambda=1$ GeV, as it also occurs for the molecular
meson contents of the  $\chi_{bJ}(3P)$ dressed states.   For $X(3872)$ meson--molecular components larger than 65 \% ($\tilde
Z_{\X} < 35\%$), both the $X_b$ and $X_{b2}$ should also exist when
$\Lambda=0.5$ GeV is used, though they would be less bound than in the $\Lambda=1$
GeV case, and for the smallest $X(3872)$ molecular component
scenarios, these states would appear now as poles in the SRS, located relatively
close to their respective thresholds. Moreover, as long as the $X_b$
and $X_{b2}$ would remain bound, 
they would present mostly a molecular nature, with quarkonium $b\bar b\,
3^3P_{1,2}$  components quite small ($\leq$ 5\% )  and less important
than in the $\Lambda=1$ GeV case, where the 
quarkonium probabilities could be larger, even of the order of 10 or
20\%. If the poles show up in the SRS, their molecular contents turn
out to be greatly reduced.

The $1^{++}$ and $2^{++}$ hidden bottom sectors were analyzed in
Ref.~\cite{Entem:2016ojz} within the quark model of
Ref.~\cite{Ortega:2010qq}. As mentioned earlier,  the $^3P_0$
phenomenological approximation  is employed in
\cite{Entem:2016ojz}  to couple quarkonium and
two-meson degrees of freedom.  As argued here, for $J^{PC}=1^{++}$ 
some repulsion from the bottomonium state
below the $B\bar B^*$ threshold is found in \cite{Entem:2016ojz}, but
however there,  it is not given  a definitive answer to the
existence or not existence of the $X_b$ state, since the results of
that work
depends critically of the strength parameter of the $^3P_0$ model
within its uncertainties. In any case, its existence is not
discarded. In the $2^{++}$ sector, an additional state,
with a mass of 10648 MeV is found in \cite{Entem:2016ojz}, and it is
pointed out that there is a  similar repulsion and attraction
from the states below ($3P$) and above ($4P$) threshold.  This state
would be just 1 or 2 MeV below the $B^*\bar B^*$ threshold, and it
could be easily accommodated within our expectations.

\section{Conclusions}
\label{sec:concl}

In this work, we have set up a scheme based on HQSS to study
quarkonium admixtures in molecular states like the $X(3872)$ or its
heavy-quark spin flavor partners, $X_2$, $X_b$ and $X_{b2}$, not
discovered yet. We have discussed how the interplay of the charmonium
components in the $X(3872)$ produces an extra attraction, and thus we
have argued that one would need less attractive meson--meson
interactions to bind the state. Such an attraction does not appear in the
$2^{++}$ sector, where one should expect instead some repulsion from
the charmonium degrees of freedom.  The $1^{++}$ bare
charmonium pole would be modified due to the $D\bar D^{(*)}$ loop
effects, and it would be moved to the complex plane acquiring also a
finite width. Despite having neglected isospin breaking terms and
working at LO in the heavy quark expansion, these effects still depend
on two unknowns LEC's.  The mass of the $X(3872)$ imposes a relation
among them, and we have considered the ratio $R_{\psi\gamma}$ of the
$X(3872)$ branching fractions into $J/\psi\gamma$ or $\psi(2S)\gamma$
to further constrain the range of variation of these two LEC's. To
that end, we have used the EFT prediction for $R_{\psi\gamma}$
obtained in Ref.~\cite{Guo:2014taa}, where meson--loop contributions
were calculated, and complemented it with the quark--loop contribution
driven by the $X(3872)\to \chi_{c1}(2P)$ transition derived here. We
have found that around a 10 -- 30\% charmonium probability (estimated
by means of the compositeness  sum-rule  of Eq.~(\ref{eq:sum-rule})) in the $X(3872)$
might explain the experimental value of the ratio $R_{\psi\gamma}$,
confirming that this ratio is not in conflict with a predominantly
molecular nature of the $X(3872)$. In turn, the dressed
$\chi_{c1}(2P)$ would have a mass and a width, which would make
plausible its identification with the $X(3940)$ resonance.

For 10 -- 30\% $c\bar c\, 2\,^3P_1$ content in
the $X(3872)$, the $X_2$ resonance destabilizes and disappears from
the spectrum, becoming either a virtual state or being located deep
into the complex plane, with decreasingly influence in the $D^{*}\bar
D^{*}$ scattering line. The crucial point here is that the
$\chi_{c2}(2P)$ state is located well below the expected mass of the
$X_2$ in the vicinity of the $D^*\bar D^*$ threshold. In sharp contrast to
what happens in the $X(3872)$ sector, where the $\chi_{c1}(2P)$ is
close  (but above) to the two meson threshold, the $\chi_{c2}(2P)$  produces a meson-meson
repulsive interaction. The $X_2(4012)$ has not been observed
yet, contrary to the HQSS expectations~\cite{Guo:2013sya}, and thus 
the study carried out here might  help to understand this fact, because
we have shown that this resonance might not be accessible to the
direct observation.

In the hidden bottom sectors and despite the changes induced by the
quarkonium admixtures, it is reasonable to expect that both $X_b$ and
$X_{b2}$ resonances might be observed in the short
future. Nevertheless, we should remind here once more than our
conclusions in the bottom sector rely on the assumption that the
contact term in the $4H$ Lagrangian and the LEC $d$, which controls
the admixtures of quarkonium and two meson configurations, are
independent of the heavy flavor. Moreover, we have also assumed that
this latter parameter does not depend on the $Q\bar Q$ radial
configuration.  Hence, it is difficult to estimate the systematic
uncertainties that affect our analysis of the $X_b$ and $X_{b2}$
resonances.  However one should bear in mind, in sharp contrast with
the $\chi_{c1}(2P)-X(3872)$ case, the bottomonium states are far
($\simeq 100$ MeV) from the $B^{(*)}\bar B^{(*)}$ thresholds. Thus, it
seems reasonable that effects due to the extra repulsion induced by
the $3P$ bottomonia in the $X_b$ and $X_{b2}$ molecular states, when
they are placed close to their respective two meson thresholds, should
not play a role as important as in the $X(3872)$.

The picture that comes out from our study turns out to be in a
remarkable agreement, at least qualitatively,  with the findings of the quark model of
Refs.~\cite{Ortega:2010qq, Entem:2016ojz}. In these works,  the $^3P_0$
phenomenological approximation  is employed  to couple quarkonium and
two-meson degrees of freedom. Thus, the $X_2$ state is not found
in \cite{Ortega:2010qq}, while the  $X(3872)$ emerges with a charmonium content similar to
that favored by our study of its radiative decays. In the $2^{++}$ hidden bottom sector, an additional state with a mass of
10648 MeV is reported in ~\cite{Entem:2016ojz}. Such state would correspond to the
$X_{b2}$,  and this mass could be  accommodated within our
predictions.  In the $1^{++}$ sector, the quark model does not provide
a definite answer about the  the existence of the
$X_b$, since the results of Ref.~\cite{Entem:2016ojz}
depends critically of the strength parameter of the $^3P_0$ model
within its uncertainties.

\appendix
\section{UV $\Lambda=500$ MeV results}
\label{sec:appendix}
In this appendix, we compile the properties of the $1^{++}$ and $2^{++}$
hidden charm (Table~\ref{tab:500charm})  and hidden bottom
(Table~\ref{tab:500bottom}) poles as a function of the mixing LEC $d$, when
an UV cutoff $\Lambda=0.5$ GeV is used to regularized the
$4H-$interactions. These results complement to those collected in
Tables~\ref{tab:dvsX}, \ref{tab:dvsX2}, \ref{tab:dvsXb} and
\ref{tab:dvsXb2}, which were obtained with $\Lambda=1$ GeV.

\begin{sidewaystable}
\centering
\begin{tabular}{cc||cc|ccc||ccc|ccc}
\multicolumn{2}{c||}{}&\multicolumn{2}{c|}{$X(3872)$}&\multicolumn{3}{c||}{$\chi_{c1}(2P)$}&\multicolumn{3}{c|}{$\chi_{c2}(2P)$}
&\multicolumn{3}{c}{$X_2$}\\
$d$ & $C_{0X}$   &  $g^{X(3872)}_{D\bar D^*}$ & $\tilde X$     &   $\left(m_{\chi_{c1}},
    \Gamma_{\chi_{c1}}\right) $ &    $g^{\chi_{c1}}_{D\bar
    D^*}$ &  $\tilde Z$      & $g^{\chi_{c2}}_{D^*\bar D^*}$  &   $\tilde
  X$ &   $\mbaredos $  & $B_{X_2}$ &    $g^{X_2}_{D^*\bar
    D^*}$ &  $\tilde X$      \\
$[{\rm fm}^{1/2}]$ & $[{\rm fm}^2]$ & $[{\rm GeV}^{-1/2}]$ & & ${\rm
    [MeV]}$ & $[{\rm GeV}^{-1/2}]$ & & $[{\rm GeV}^{-1/2}]$  
  & &  ${\rm [MeV]}$ & ${\rm [MeV]}$& $[{\rm GeV}^{-1/2}]$ &  \\\hline
0 & $-1.94$ &  1.05  &        1        &    $(3906, 0)$  &         0
&       1         &     0 &       0 &     3927.2 ~    &   $-4.8$  &
 1.10  &        1                  \\
0.1    & $-1.88$  &  1.04  &        0.98~     &    ~(3906.7, 1.5)~~ &
$0.06 + 0.13\,i$~ & $0.99 + 0.01\,i$ ~&     0.66 &   0.0  &   3927.7
&   $-3.9$    &         1.00  &        1.00            \\
0.2 & $-1.71$  &  1.02     &     0.93     &    (3908.8, 6.3)  &  $0.13
+ 0.24\,i$  & $0.96 + 0.05\,i$  &     1.26 &   0.02  &  3928.9~     &
$-1.7$   &          0.73    &     0.99  \\
0.3  &  $-1.42$  &  0.98    &      0.86     &    (3912.3, 15.6)   &
$0.21 + 0.33\,i$ & $0.92 + 0.11\,i$  &     1.77  &  0.03  &
  3930.9     &   $-0.0$ at SRS  &   $-0.08\,i$   &      $> 1$
  \\
0.4 &       $-1.02$  &  0.93   &         0.78     &    (3917.5, 31.9)
&  $0.30 + 0.40\,i$ &  $0.87 + 0.21\,i$ &     2.16 &    0.05  &
3933.2     &   $-8.3$ at SRS &    $-0.69\,i$ &   $> 1$ \\
0.5  &      $-0.50$  &  0.87   &       0.69     &    (3925.4, 61.2) &
$ 0.41 + 0.45\,i$ & $0.77 + 0.37\,i$  &     2.44  &  0.06 &   3935.7
&   $-10.6 -\frac{102.6\,i}{2}$~   & $0.05 + 0.49\,i$~ &  $0.53 + 0.01\,i$  \\
 $d^{\rm crit}$ & 0.0   &  0.83   &       0.62     &    (3938.6, 102.6)
& $ 0.51 + 0.51\,i$  & $0.57 + 0.56\,i$  &     2.60  &  0.07 &
3937.8     &    $27.7 - \frac{181.9\,i}{2}$ &   $0.26 + 0.55\,i$  &
$0.77 - 0.25\,i$\\
0.7 &       0.88  &  0.77   &       0.53     &    (3809.7,0) at SRS &
0.38   &       1.36          &     2.74  &   0.08  &   3940.7     &
$107.9 - \frac{187.8\,i}{2}$ &  $0.37 + 0.57\,i$  & $0.94 - 0.12\,i$ \\\hline
\end{tabular}
\caption{Properties of the $1^{++}$ and $2^{++}$
hidden charm poles as a function of $d$.  We solve  Eq.~(\ref{eq:pole-position})
with $\Lambda= 0.5$ GeV and for each value of $d$, $C_{0X}$ is determined from Eq.~(\ref{eq:defc0x}). The position of the $X(3872)$ is fixed at
  $M_X=3871.69$ MeV in the FRS. The $\chi_{c1}(2P)$ pole is located
  in the SRS, while the position of the dressed $\chi_{c2}(2P)$ is
fixed at $m_{\chi_{c2}}^{\rm exp}=3927.2$ MeV in the FRS. Finally,
$B_{X_2}= M_{X_2}-2 M_{D^*}-i\frac{\Gamma_{X_2}}{2} $ and $d^{\rm \,crit}(\Lambda=0.5\,{\rm GeV})= 
  \sqrt{\frac{M_X-\mbareuno}{G_{\rm QM}^I(M_X)}}=0.580$ fm$^{1/2}$.}\label{tab:500charm}
\end{sidewaystable}
\begin{sidewaystable}
\centering
\begin{tabular}{cc||ccc|ccc||ccc|ccc}
\multicolumn{2}{c||}{} & \multicolumn{3}{c|}{$\chi_{b1}(3P)$}&\multicolumn{3}{c||}{$X_b$}&\multicolumn{3}{c|}{$\chi_{b2}(3P)$}&\multicolumn{3}{c}{$X_{b2}$}\\
$d$  & $\tilde X_{X(3872)}$ &   $g^{\chi_{b1}}_{B\bar B^*}$ &   $\tilde
  X$ & $\mbareunob $  & $B_{X_b}$ & $g^{X_b}_{B\bar
    B^*}$ &  $\tilde X$  & $g^{\chi_{b2}}_{B^*\bar B^*}$ &   $\tilde
  X$ & $\mbaredosb $  & $B_{X_{b2}}$ & $g^{X_{b2}}_{B^*\bar
    B^*}$ &  $\tilde X$ \\
$[{\rm fm}^{1/2}]$ & &  $[{\rm GeV}^{-1/2}]$ 
  & &  ${\rm [MeV]}$ & ${\rm [MeV]}$& $[{\rm GeV}^{-1/2}]$ & & $[{\rm
      GeV}^{-1/2}]$  
  & &  ${\rm [MeV]}$ & ${\rm [MeV]}$ & $[{\rm GeV}^{-1/2}]$\\\hline
0 & 1  &      0   &    0  &     10512.1~  &    $-24.2$
&     2.43     &      1            &    0 &     0 &     10522.1~    &
$-24.2$  &           2.44  &    1\\      
0.1& 0.98 &  2.60  &   0.01  &  10512.8   &    $-22.3$ &
2.21   &        0.99  &    4.91   &  0.0  &   10522.6    &
$-22.6$  &           2.24 &         1.0 \\
0.2 & 0.93      &    4.85  &  0.03 &   10514.7   &    $-16.9$   &
1.65    &       0.97          &   9.32  &  0.02 &  10523.8    &
$-17.7$ &              1.74  &        0.99\\
0.3 & 0.86    & 6.56  &  0.06  &  10517.4   &    $-8.9$   &
0.96   &        0.95          &  13.04  &  0.03  & 10525.8    &
$-10.3$    &         1.08     &     0.97\\
0.4 & 0.78   &   7.72 &   0.08  &  10520.4   &    $-0.8$  &
0.30    &       0.96          &  15.92  &   0.04  &  10528.2    &
$-2.1$    &         0.43  &        0.97\\
0.5 & 0.69   &  8.42 &   0.10  &  10523.6   &
$3.2 - \frac{48.2\,i}{2}$~~  & $0.02 + 0.20\,i$~ & $0.57 + 0.07\,i$~     &
17.99   &  0.06  & 10530.7    &  $-13.7 - \frac{30.3\,i}{2}$~ &
$-0.06 + 0.20\,i$~ & $ 0.55 + 0.52\,i$ \\
$d^{\rm crit}$ & 0.62 &    8.72  &  0.10  &  10526.0   &  $41.6 -
\frac{63.6\, i}{2}$ & $0.14 + 0.21\, i$  & $0.82 - 0.12\, i$      &
19.13  &  0.06 &  10532.8    &   $35.6 - \frac{71.2\, i}{2} $ &
$0.12 + 0.21\,i$  & $0.86 - 0.09\, i$\\
0.7 &  0.53 &  8.87 &   0.11 &   10529.1   &  $81.5 -
\frac{43.3\,i}{2}$ &  $0.19 + 0.14\,i$ & $0.90 - 0.06\,i$      
&  20.13  &  0.07  & 10535.7    &   $78.1 - \frac{46.9\,i}{2}$ &
$0.18 + 0.15\, i$ & $0.93 - 0.04\, i$ \\\hline
\end{tabular}
\caption{Properties of the $1^{++}$ and $2^{++}$
hidden bottom poles as a function of $d$.  We solve  Eq.~(\ref{eq:pole-position})
with $\Lambda= 0.5$ GeV and $C_{0X}(d)$, determined from
Eq.~(\ref{eq:defc0x}),  can be found  in Table~\ref{tab:500charm}. The position of the dressed $\chi_{b1}(3P)$ and
$\chi_{b2}(3P)$ are fixed at 10512.1 and 10522.1 MeV in the FRS. The positions of the $X_b$ and $X_{b2}$ poles are determined by
$B_{X_b}= M_{X_b}-M_B- M_{B^*}-i\frac{\Gamma_{X_b}}{2} $ and
$B_{X_{b2}}= M_{X_{b2}}-2 M_{B^*}-i\frac{\Gamma_{X_{b2}}}{2} $,
  respectively. The LEC $d^{\rm \,crit}(\Lambda=0.5\,{\rm
    GeV})=0.580$ fm$^{1/2}$ reproduces the mass of the
  $X(3872)$ with $C_{0X}=0$ and $ \Lambda= 0.5$ GeV. Note that we also
  give the $X(3872)$ meson-molecular
probabilities  ($\tilde X_{X(3872)}$) for each value of $d$.} \label{tab:500bottom}
\end{sidewaystable}

\begin{acknowledgments}
We  would like to thank M. Albaladejo, D.R. Entem,
P. Fern\'andez-Soler,  F.-K. Guo and J.A. Oller for
enlightening comments.  This research
 has been supported by the Spanish Ministerio de Econom\'\i a y
 Competitividad and European FEDER funds under the contracts
 FIS2014-51948-C2-1-P,  FIS2014-57026-REDT and SEV-2014-0398, by
 Generalitat Valenciana under contract PROMETEOII/2014/0068 and by
 TUBITAK under contract 114F234.
\end{acknowledgments}

\bibliography{charmonium}

\end{document}